\documentclass[fleqn,usenatbib]{mnras}

\usepackage{newtxtext,newtxmath}
\usepackage[T1]{fontenc}
\usepackage{ae,aecompl}
\usepackage{array}
\usepackage{graphicx}	% Including figure files
\usepackage[nice]{nicefrac}
\usepackage{bm}
\usepackage{textcomp}
\usepackage{hyperref}
\usepackage{natbib}
\usepackage{amsmath}
\usepackage{graphicx}
\usepackage[colorinlistoftodos]{todonotes}
\usepackage{xcolor}
\usepackage{multirow}
\usepackage{todonotes}
\usepackage{float}
\usepackage{booktabs}
\usepackage{subfig}
\usepackage{lineno}
\usepackage{seqsplit}
\usepackage{verbatim}
\usepackage{ulem}

\usepackage{color}
\definecolor{darkblue}{rgb}{0.,0.,0.4}
\definecolor{darkred}{rgb}{0.5,0.,0.}
\hypersetup{urlcolor=darkred, citecolor=blue, linkcolor=blue}

\definecolor{ForestGreen}{RGB}{34, 139, 34}

\newcommand{\bb}[1]{\mathbb{#1}}
\newcommand{\bs}[1]{\boldsymbol{#1}}
\newcommand{\mc}[1]{\mathcal{#1}}
\newcommand{\tm}[1]{\textrm{#1}}
\newcommand{\sans}[1]{\bm{\mathsf{#1}}}

\title[Inference with GP and MOPED for WL]{Parameter Inference for Weak Lensing using Gaussian Processes and MOPED}

\author[A. Mootoovaloo et al.]{Arrykrishna Mootoovaloo$^{1}$\thanks{E-mail: a.mootoovaloo17@imperial.ac.uk}, Alan F. Heavens$^{1}$, Andrew H. Jaffe$^{1}$ and Florent Leclercq$^{1}$\\
$^{1}$ICIC, Astrophysics, Imperial College, Blackett Laboratory, Prince Consort Road, London SW7 2AZ, UK}

\date{Accepted 2020 July 10. Received 2020 July 03; in original form 2020 May 19}
\pubyear{2020}

\begin{document}
\label{firstpage}
\pagerange{\pageref{firstpage}--\pageref{lastpage}}
\maketitle
%%%%%%%%%%%%%%%%%%%%%%%%%%%%%%%%%%%%%%%%%%%%%%%%%%

\begin{abstract}
\label{firstpage}

In this paper, we propose a Gaussian Process (GP) emulator for the calculation both of tomographic weak lensing band-powers, and of coefficients of summary data massively compressed with the MOPED algorithm.  In the former case cosmological parameter inference is accelerated by a factor of $\sim 10$-$30$ compared with Boltzmann solver CLASS applied to KiDS-450 weak lensing data.  Much larger gains of order $10^3$ will come with future data, and MOPED with GPs will be fast enough to permit the  Limber approximation to be dropped, with acceleration in this case of $\sim 10^5$.  A potential advantage of GPs is that an error on the emulated function can be computed and this uncertainty incorporated into the likelihood. However, it is known that the GP error can be unreliable when applied to deterministic functions, and we find, using the Kullback-Leibler divergence between the emulator and CLASS likelihoods, and from the uncertainties on the parameters, that agreement is better when the GP uncertainty is not used. In future, weak lensing surveys such as Euclid, and the Legacy Survey of Space and Time (LSST), will have up to $\sim 10^{4}$ summary statistics, and inference will be correspondingly more challenging.  However, since the speed of MOPED is determined  not the number of summary data, but by the number of parameters, MOPED analysis scales almost perfectly, provided that a fast way to compute the theoretical MOPED coefficients is available.  The GP provides such a fast mechanism.
\end{abstract}

\section{Introduction}

With the continuous advancement of technology, our understanding of the Universe has progressively improved, starting with the radiation of the cosmic microwave background (CMB) from experiments such as COBE \citep{1992ApJ...396L...1S, 2001PhRvL..86.3475J}, WMAP \citep{2003ApJS..148..175S} and Planck \citep{2014A&A...571A..16P, 2016A&A...594A..13P} to ongoing observations of supernovae as standard candles \citep{2014A&A...568A..22B} and large-scale structure probes \citep{2014MNRAS.441...24A}. While these experiments place the $\Lambda$CDM model on a firm footing, there is still the need to understand better dark matter and the evolution of dark energy. A powerful probe of the geometry of the Universe is cosmic shear, the weak lensing effect observed as a result of bending of light between the observer and background galaxies due to intervening large-scale structure. Cosmic shear is proving to be a powerful method for measuring the temporal and spatial properties of dark species statistically over a large sample of sources \citep{2017MNRAS.471.4412K}.

However, one possible bottleneck in the era of massive surveys lies in the forward computation of two-point statistics from cosmological parameters, either correlation functions or power spectra \citep{2015RPPh...78h6901K}. The complexity of the problem is exacerbated when performing weak lensing analysis in $n$ tomographic redshift slices, for which $n(n+1)/2$ auto- and cross-correlations are needed for each angular scale probed. If intrinsic alignments \citep{2000MNRAS.319..649H, 2004PhRvD..70f3526H} are included, we require an additional calculation of $n(n+1)$ power spectra. In general, running this full forward model is computationally expensive. The problem is even worse in very large simulation settings. As an example, \cite{2009ApJ...705..156H} argued that a naive analysis to obtain cosmological constraints from future suveys such as LSST \citep{2002SPIE.4836..111S, 2017arXiv170804058L} from weak lensing shear spectra will take a 2048 processor machine 20 years. An alternative approach is to replace the simulator (the full forward model) by an approximate mathematical function, often referred to as a metamodel, surrogate model, or emulator.

Various types of emulators have been designed for different purposes in cosmology. As an example, PICO \citep{2007ApJ...654....2F} was developed to accelerate parameter estimation for the CMB. The underlying idea is to use an order-$p$ polynomial and a clustering method to interpolate between power spectra generated at specific points in parameter space. On the other hand, neural networks have also been been used as emulators for generating power spectra. \cite{2007MNRAS.376L..11A} used a 3-layer multilayer perceptron (MLP) algorithm to construct an emulator to accelerate the calculation of the CMB power spectra. Neural networks learn the non-linearities between the input space and output space by optimizing the weights associated with each neuron. In the same spirit, \cite{2018MNRAS.475.1213S} developed a 2-layer MLP algorithm to emulate the 21cm power spectrum. Similarly, \cite{2012MNRAS.424.1409A, 2014MNRAS.439.2102A} designed a neural network emulator for non-linear matter power spectrum interpolation based on 6 cosmological parameters. \cite{2018PhRvD..97h3004C} also used neural networks to find functionals of the data that maximise the Fisher information matrix. \cite{2020MNRAS.491.2655M} devised a neural network scheme, coupled to an MCMC, to accelerate cosmological parameter inference. In an alternative approach, \cite{2007PhRvD..76h3503H} developed a statistical framework using Gaussian Processes to emulate matter power spectrum and \cite{2011ApJ...728..137S} used a similar method to emulate cosmic microwave background temperature power spectrum. An analogous approach was used and extended in the Coyote Universe collaboration \citep{2009ApJ...705..156H, 2010ApJ...715..104H, 2010ApJ...713.1322L, 2014ApJ...780..111H}. Recently, \cite{2019JCAP...02..050B} and \cite{2019JCAP...02..031R} developed a Gaussian Process emulator for Lyman-$\alpha$ forest simulation. \cite{2017ApJ...848...23K} designed a Gaussian Process emulator for the 21cm power spectrum.  

An alternative to predicting the signal using Machine Learning techniques is to predict the likelihood. Instead of building emulators at the level of the theory/data, another option is to construct a likelihood regressor. \cite{2007arXiv0712.0194F} extended the PICO formalism to fit a likelihood function. \cite{2018PhRvD..98f3511L} developed the BOLFI algorithm \citep{2015arXiv150103291G} to construct a likelihood regressor which fits two cosmological parameters $w_{0}$ and $\Omega_{\textrm{m}}$ for the JLA dataset \citep{2014A&A...568A..22B}. In particular, in the BOLFI approach, the uncertainty from the Gaussian Process appears in the acquisition function, which is used to choose the next design point where the simulation will be run. However, the surrogate model itself is not a perfect reconstruction of the true likelihood function. Instead, if the emulator were built at the level of the theoretical model, the GP uncertainty can be propagated in the full forward model. In the same spectrum, \cite{2018MNRAS.477.2874A} developed a density-estimation likelihood-free method and argued that it requires less tuning compared to traditional approaches such as Approximate Bayesian Computation, ABC. \cite{2019MNRAS.490.4237L} also developed a likelihood-free approach to infer power spectrum and cosmological parameters from forward simulations only.

Each type of emulator (polynomial, neural network and Gaussian Processes) has its own advantages and disadvantages. In the case of polynomial regression, one has to specify the basis functions prior to performing interpolation. However, it provides a full predictive distribution of the function at a test point. On the other hand, neural network regression requires much empirical work, for example, choosing an appropriate optimiser, setting the learning rate of the optimiser and choosing the number of layers and neurons. The training stage is computationally very challenging if we have tens or hundreds of functions to interpolate, as in our case, but it can be done \citep{2020MNRAS.491.2655M}. The outputs from a neural network are point estimates, unless we consider Bayesian neural networks which attempt to get a full posterior distribution on the weights of the neural network using variational inference methods, but this is also challenging \citep{2017arXiv170304977K}. 

In this paper, we develop a Gaussian Process emulator for speeding cosmological parameter inference for weak lensing cosmology whilst keeping the number of calls to the full solver to the order of a few tens of thousands only, to avoid an expensive and potentially unstable Cholesky decomposition.  In other contexts, where the generation of training point data may be done through expensive full numerical simulations, the number may need to be further restricted. Despite this restriction, an important and interesting result is that the posterior densities between the emulator and the full solver are comparable. On the other hand, in the case of full N-body simulations, one might be restricted to only a few hundred forward simulations, and the mechanism (MOPED + GP) proposed in this work (see discussion below) will enable us to develop promising inference schemes for this scenario. A useful advantage of GPs is that we also get a theoretical predictive uncertainty. However, note that the latter depends on the Gaussian Process model, including the kernel. Building such a framework is a challenging task. It is well-known that non-parametric regression methods such as Gaussian Processes suffer from the so-called `{curse of dimensionality}' \citep{geenens2011curse} as a result of the training data being sparse in high dimensional spaces. The only way to obtain accurate interpolation is to add information to our model and this can be achieved by adding more and more training points. However, this involves a penalty, not only in terms of computations but also storage. Moreover, from the cosmology perspective, drawing inferences from current weak lensing data is a demanding task, for we have to correctly account for sources of systematic error which add new effective parameters to our model. In addition to this, given the current status of the weak lensing field with relatively few data points at low signal-to-noise, cosmological parameters are not very well-constrained compared to (e.g.) the CMB. As a result, our training points have to be distributed over a larger volume in parameter space.

Fortunately, sophisticated techniques such as Latin Hypercube (LH) sampling \citep{mckay1979comparison}, with the appropriate transformations at the level of both the input covariates and the response, can simplify the problem.  Re-casting the full problem as a hierarchical model enables us to leverage the predictive uncertainty of the reconstructed probabilistic functions, thus propagating the uncertainty consistently in the statistical framework. This is in principle a major advantage of Gaussian Processes, but the GP error is not always a reliable measure when applied to deterministic functions \citep{2020arXiv200110965K, 2020arXiv200201381W}, and we find that the posteriors are more accurate if it is not included (see \S\ref{sec:results} for further details).

A powerful application of the GP emulator is to combine it with the extreme data compression algorithm MOPED \citep{2000MNRAS.317..965H} to further speed up parameter inference. MOPED is essentially a lossless data compression technique which, irrespective of the size of the original data, compresses the latter to the number of parameters in the model. One can simply replace the theoretical prediction by the surrogate model in the MOPED likelihood function and use the MOPED vectors to compute the MOPED coefficients. However, in order to achieve the full acceleration that MOPED allows, a fast mechanism to generate the expected coefficients themselves is needed, and GPs can do this.

In this study, as a first application, the Gaussian Process emulator is built firstly for tomographic weak lensing band power spectra, and secondly for MOPED coefficients. With a well-designed emulator, it is possible to obtain reliable marginalised posterior densities of cosmological and nuisance parameters with a few hundred forward simulations only. We find that for a small number of training points, the emulator yields posterior distributions close to the true posterior whilst still maintaining a decent speed for each likelihood evaluation.  When combined with MOPED data compression, the GP allows for very large acceleration of analysis of large datasets, which may otherwise be prohibitively expensive.

The paper is organised as follows: in \S\ref{sec:weak_lensing} we provide a brief overview on weak lensing theory and in \S\ref{sec:gp}, we touch upon Gaussian Processes before systematically going through the main steps taken to build the emulator in \S\ref{sec:emulator}. In \S\ref{sec:data_compression}, we discuss how the MOPED algorithm can be used alongside the emulator and we present our results in \S\ref{sec:results} before concluding in \S\ref{sec:conclusions}. Throughout, we will assume a flat universe. 

\begin{figure}
\noindent \begin{centering}
\includegraphics[width=0.45\textwidth]{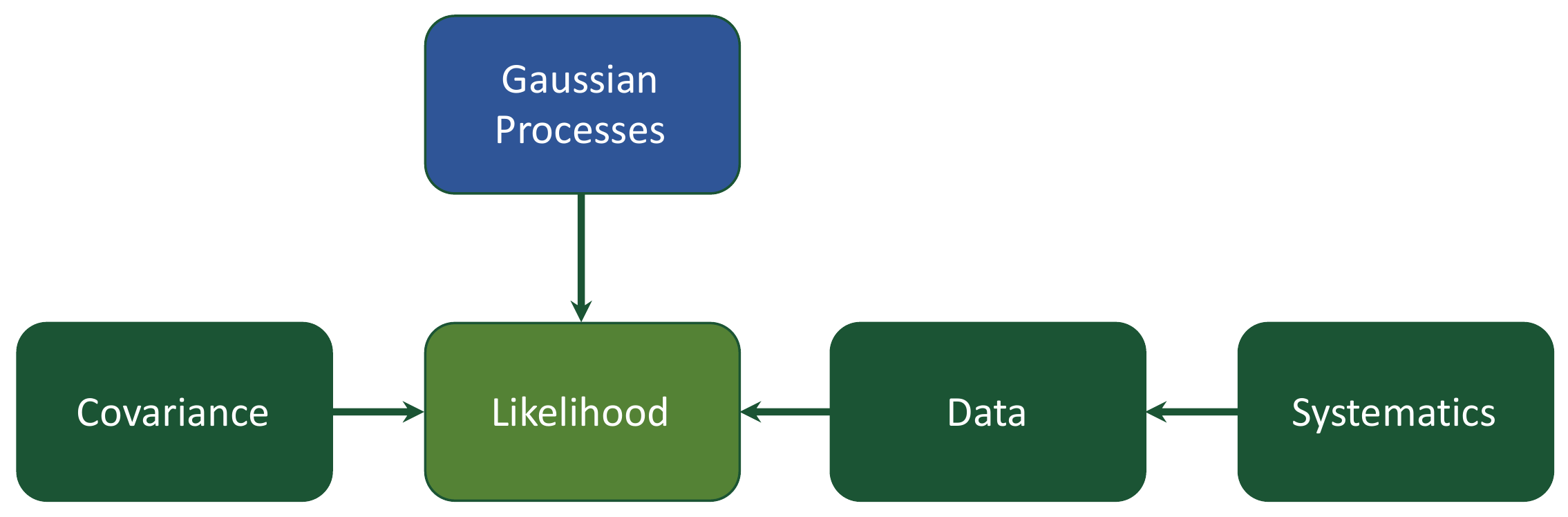}
\par\end{centering}
\caption{\label{fig:idea-pipeline}A diagrammatic form of the core principle in this work. We substitute the most expensive part of the pipeline by surrogate models (Gaussian Processes) built at the level of the band powers. The other blocks in the inference procedure, for example, for the computations related to the nuisance parameters, are unaltered.}   
\end{figure}
\section{Weak Lensing}
\label{sec:weak_lensing}

Gravitational lensing is the bending of light as it propagates through the inhomogeneous Universe, which leads to a coherent distortion of galaxy images. In particular, weak lensing can only be studied in a statistical sense since the distortion is small and requires averaging over a large sample of galaxies as a result of broad distribution of intrinsic ellipticities. Weak lensing has the advantage of probing the distribution of matter and not of biased tracers such as galaxies which is hard to predict. In essence, we would ideally want to perform a full 3d statistical analysis but much of the cosmological information can still be retained by tomography, where objects are separated by redshift. We refer the reader to common literature on weak gravitational lensing \citep{2001PhR...340..291B, 2015RPPh...78h6901K} for further details on these techniques. 

In a weak lensing analysis, observables include galaxy positions, photometric redshifts and shapes. The latter is given in terms of the ellipticity components, $\epsilon_{1}$ and $\epsilon_{2}$. In particular, this observed ellipticity, $\bs{\epsilon}=\epsilon_{1}+i\epsilon_{2}$ is related to the unlensed intrinsic ellipticity, $\bs{\epsilon}_{0}$ via the shear field such that
\begin{equation}
\bs{\epsilon} \simeq \bs{\epsilon}_{0} + \bs{\gamma}.
\end{equation}
\noindent The observed ellipticities are binned into pixels and redshift bin, $i$. An estimate of the shear field, $\bs{\hat{\gamma}}_{i}$ is obtained by averaging the ellipticities in each pixel. This (complex) shear field can be expanded in terms of spin-weighted spherical harmonics, $_{s}Y_{\ell m}$ \citep{2000PhRvD..62d3007H, 2005PhRvD..72b3516C}

\begin{equation}
\gamma_{1}(\bs{r})\pm i\gamma_{2}(\bs{r})=\sqrt{\dfrac{2}{\pi}}\sum_{\ell m}\int \tm{d}k\,k^{2}\gamma_{\ell m}(k)_{\pm 2}Y_{\ell m}(\hat{\bs{n}})j_{\ell}(kr),
\end{equation}
\noindent where $j_{\ell}$ is a spherical Bessel function, $k$ is a radial wavenumber and $\ell$ is a positive integer while $m=-\ell,\ldots\ell$. The coefficients $\gamma_{\ell m}$ are related to the transform of the lensing potential, $\phi(\bs{r})$ by
\begin{equation}
\gamma_{\ell m} =\dfrac{1}{2}\sqrt{\dfrac{(\ell+2)!}{(\ell-2)!}}\phi_{\ell m}(k).
\end{equation}

\noindent Similarly, the expansion coefficients for the convergence field, $\kappa$, is
\begin{equation}
\kappa_{\ell m} = -\dfrac{\ell(\ell+1)}{2}\phi_{\ell m} (k).
\end{equation} 

The shear field can be decomposed into E and B modes corresponding to the curl-free and divergence-free components. In particular, the convergence field, $\kappa^{\tm{E}}$ contains most of the cosmological information since $\kappa^{\tm{B}}$ is negligible in the absence of systematics \citep{2005PhRvD..72b3516C}. Under this condition, the E-mode lensing power spectrum between tomographic bins $i$ and $j$ is equal to the convergence power spectrum, that is, $C_{\ell,\,ij}^{\tm{EE}}=C_{\ell,\,ij}^{\kappa\kappa}$ and is given, in the Limber approximation \citep{1953ApJ...117..134L, 2008PhRvD..78l3506L} by

\begin{equation}
C_{\ell,\,ij}^{\tm{EE}}=\int_{0}^{\chi_{H}}\tm{d}\chi\,\dfrac{w_{i}(\chi)w_{j}(\chi)}{\chi^{2}}\,P_{\delta}\left(k=\dfrac{\ell+\nicefrac{1}{2}}{\chi};\,\chi\right),
\end{equation}

\noindent where $\chi$ is the comoving radial distance and $\chi_{H}$ is the comoving distance to the horizon. Without the Limber approximation, the integrals can be slow to compute, although faster methods are being developed \citep{2019arXiv191111947F}. Crucially, the tomographic convergence power spectrum is sensitive to the background geometry and the growth of structure. It depends on the the three-dimensional matter power spectrum, $P_{\delta}(k;\chi)$ which is a function of redshift \citep{2013PhR...530...87W}. The weight function $w_{i}$ for a flat universe is

\begin{equation}
\label{weight_function}
w_{i}(\chi)=\dfrac{3\Omega_{\tm{m}}H_{0}^{2}}{2c^{2}}\chi(1+z)\int_{\chi}^{\chi_{H}}\tm{d}\chi'n_{i}(\chi')\left(\dfrac{\chi'-\chi}{\chi'}\right),
\end{equation}

\noindent which depends on the lensing kernel. $\Omega_{\tm{m}}$ is the present matter density, $H_{0}$ is the Hubble constant and $c$ is the speed of light. An important quantity is the redshift distribution, $n_{i}\left(z\right)dz=n_{i}\left(\chi\right)d\chi$ which is normalised such that 

\begin{equation}
\int n_{i}(\chi)d\chi=1.
\end{equation}

For a weak lensing survey, the data vector consists of the measured shear per pixel for each redshift bin. At this point, in order to extract the shear power spectrum, one can either take a quadratic estimator approach using maximum-likelihood technique \citep{1998PhRvD..57.2117B} or employ a pseudo-$C(\ell)$ approach \citep{2007ApJS..170..288H}. Alternatively, one can also build a full Bayesian hierarchical model, to infer the full shear power power spectrum \citep{2016MNRAS.455.4452A, 2017MNRAS.466.3272A}. Here, we focus on the tomographic band power spectra, as determined by \cite{2017MNRAS.471.4412K}.

\begin{figure*}
\noindent \begin{centering}
\includegraphics[width=0.8\textwidth]{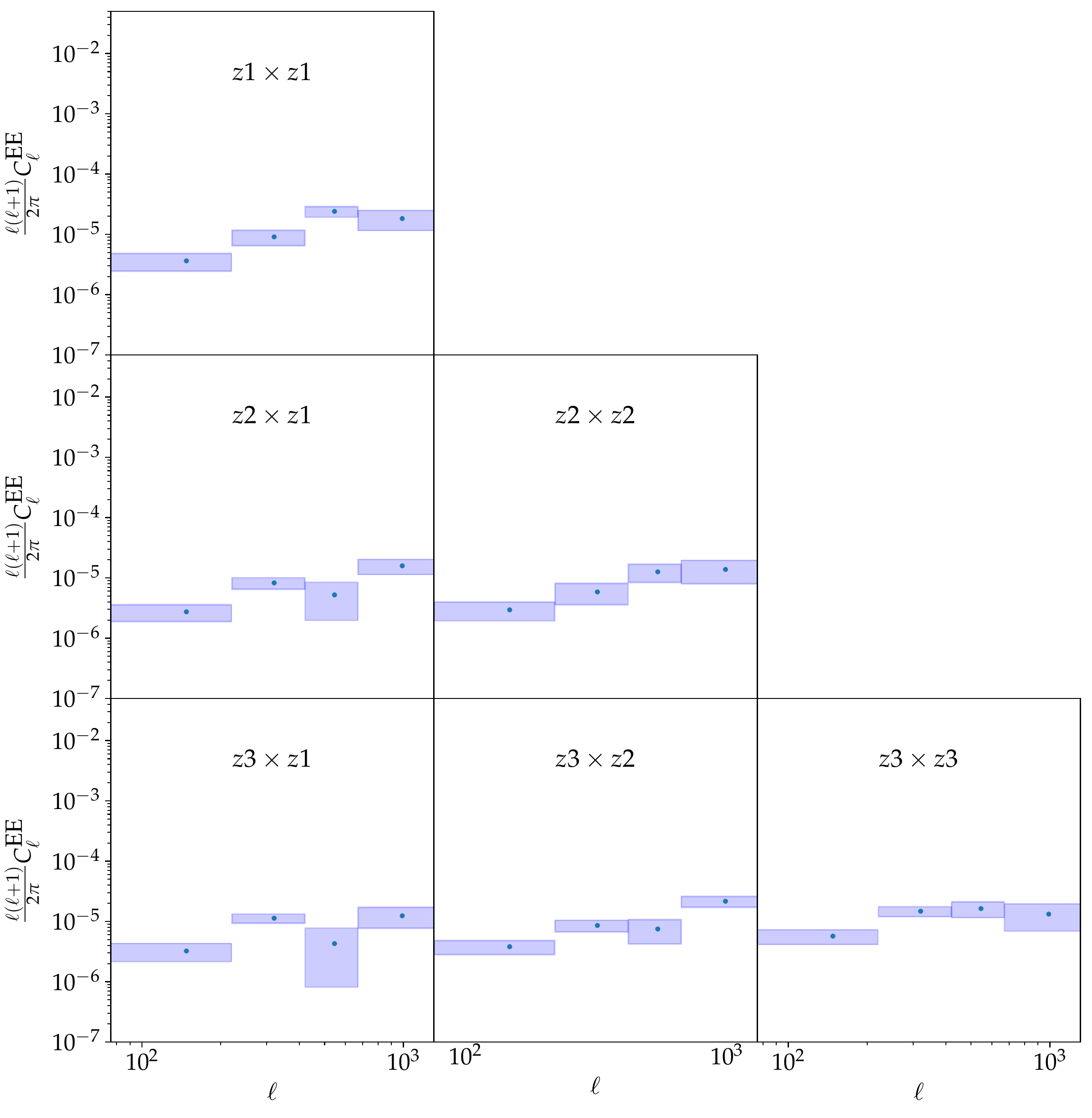}
\par\end{centering}
\caption{\label{fig:kids-data}The E-mode band powers (data) used in our inference scheme, similar to the KiDS-450 analysis \citep{2017MNRAS.471.4412K}. The $\ell-$ranges are as follows: $76\leq\ell<220$,  $221\leq\ell<420$, $421\leq\ell<670$ and $671\leq\ell<1310$. In particular, the auto-correlation band powers are along the main diagonal ($z1 \times z1$, $z2 \times z2$ and $z3 \times z3$) for the 3 redshift bins $0.10<z_{1}\leq0.30$, $0.30<z_{2}\leq0.60$ and $0.60<z_{3}\leq0.90$. The off-diagonal blocks show the unique cross-correlation band powers. The blue shaded regions indicate the $1\sigma$ level errors from the covariance matrix.}
\end{figure*}

\subsection{Astrophysical Systematics}
\label{sec:astrophysical_systematics}

Coupled to the E-mode power spectrum are various systematics which we should consider. For example, baryon feedback results in altering the power in high $k$. Although feedback is not fully understood, it is often parameterized through the bias function, $b^{2}(k,z)$ such that the modified power spectrum is

\begin{equation}
P_{\delta}^{\tm{mod}}\left(k,z\right)=b^{2}\left(k,z\right)\,P_{\delta}\left(k,z\right).
\end{equation}

As an example, for the KiDS-450 analysis, the following fitting formula from \cite{2011MNRAS.415.3649V} was used

\begin{equation}
b^{2}\left(k,z\right)=1-A_{\tm{bary}}\left[A_{z}e^{\left(B_{z}x-C_{z}\right)^{3}}-D_{z}x e^{E_{z}x}\right],
\end{equation}

\noindent where $x=\textrm{log}_{10}(k/1\textrm{ Mpc}^{-1})$ and the other parameters $A_{z},\,B_{z},\,C_{z},\,D_{z}$ and $E_{z}$ depend on the scale factor $a$. Moreover, we must account for intrinsic alignment effects which give rise to a preferred ellipticity orientation. The total lensing power spectrum between two redshift slices is a linear combination of the gravitational lensing (EE), intrinsic alignment (II) and interference (GI) power spectra. Specifically, the II effect is due to correlation of ellipticities in the local environment and contributes positively towards the total lensing spectrum. The second effect, GI, is due to correlation between tidally-stretched foreground galaxies and the shear of background galaxies. The GI term subtracts from the total lensing spectrum. We model the power spectrum, following \cite{2017MNRAS.471.4412K}, as

\begin{equation}
\label{eq:cl_tot}
C_{\ell,\,ij}^{\tm{tot}}=C_{\ell,\,ij}^{\tm{EE}}+A_{\tm{IA}}^{2}C_{\ell,\,ij}^{\tm{II}}-A_{\tm{IA}}C_{\ell,\,ij}^{\tm{GI}},
\end{equation}

\noindent where the II power spectrum, $C_{\ell,\,ij}^{\tm{II}}$ and the GI power spectrum, $C_{\ell,\,ij}^{\tm{GI}}$ respectively are

\begin{equation}
C_{\ell,\,ij}^{\tm{II}}=\int_{0}^{\chi_{H}}\tm{d}\chi\,\dfrac{w_{i}(\chi)w_{j}(\chi)}{\chi^{2}}\,P_{\delta}\left(k=\dfrac{\ell+0.5}{\chi};\,\chi\right)\,F^{2}\left(\chi\right),
\end{equation}

\begin{equation}
\begin{split}
C_{\ell,\,ij}^{\tm{GI}}=\int_{0}^{\chi_{H}}\tm{d}\chi\,\dfrac{w_{i}(\chi)n_{j}(\chi)+w_{j}(\chi)n_{i}(\chi)}{\chi^{2}}\\
P_{\delta}\left(k=\dfrac{\ell+0.5}{\chi};\,\chi\right)\,F(\chi),
\end{split}
\end{equation}

\begin{equation}
F(\chi)=C_{1}\rho_{\tm{crit}}\dfrac{\Omega_{m}}{D_{+}(\chi)},
\end{equation}

\noindent and $A_{\tm{IA}}$ is a free parameter to be inferred during sampling. This allows for the flexibility of rescaling the otherwise fixed normalisation value, $C_{1}=5\times10^{-14}\;h^{-2}\,\tm{M}_{\odot}^{-1}\,\tm{Mpc}^{3}$. $\rho_{\tm{crit}}$ is the critical density of the Universe while $D_{+}(\chi)$ refers to the linear growth factor normalized to unity today. 
\section{Gaussian Processes}
\label{sec:gp}

In this section, we provide a general outline of the basic concepts behind the Gaussian Process, which is at the core of our cosmological parameter inference scheme. We refer the reader to \cite{2006gpml.book.....R} for further details.

Our goal is to design an emulator, which takes as input a set of cosmological parameters, $\bs{\theta}$, and outputs summary statistics, $\bs{y}$, for example, power spectrum or band powers. We consider our regression problem to be in the form,

\begin{equation}
\bs{y}=\bs{f}+\bs{\epsilon},
\end{equation}

\noindent where $f$ is a multivariate Gaussian when evaluated at an arbitrary number of points, and $\epsilon$ is an error.  The distribution of $f$ is informed by a set of $N_{\tm{train}}$ training points in parameter space, and has a Gaussian prior distribution, $f\sim\mc{N}(0,k_{pq})$ with a covariance $k_{pq}\equiv k(\bs{\theta}_{p},\,\bs{\theta}_{q})$.

The kernel function $k$ which encapsulates the correlation between points in the parameter space is used to model the smoothness of the function. We choose the automatic relevance determination (ARD) kernel, also referred to as the Squared Exponential (SE) kernel, Radial Basis Function (RBF) or simply the Gaussian kernel:

\begin{figure}
\noindent \begin{centering}
\includegraphics[width=0.45\textwidth]{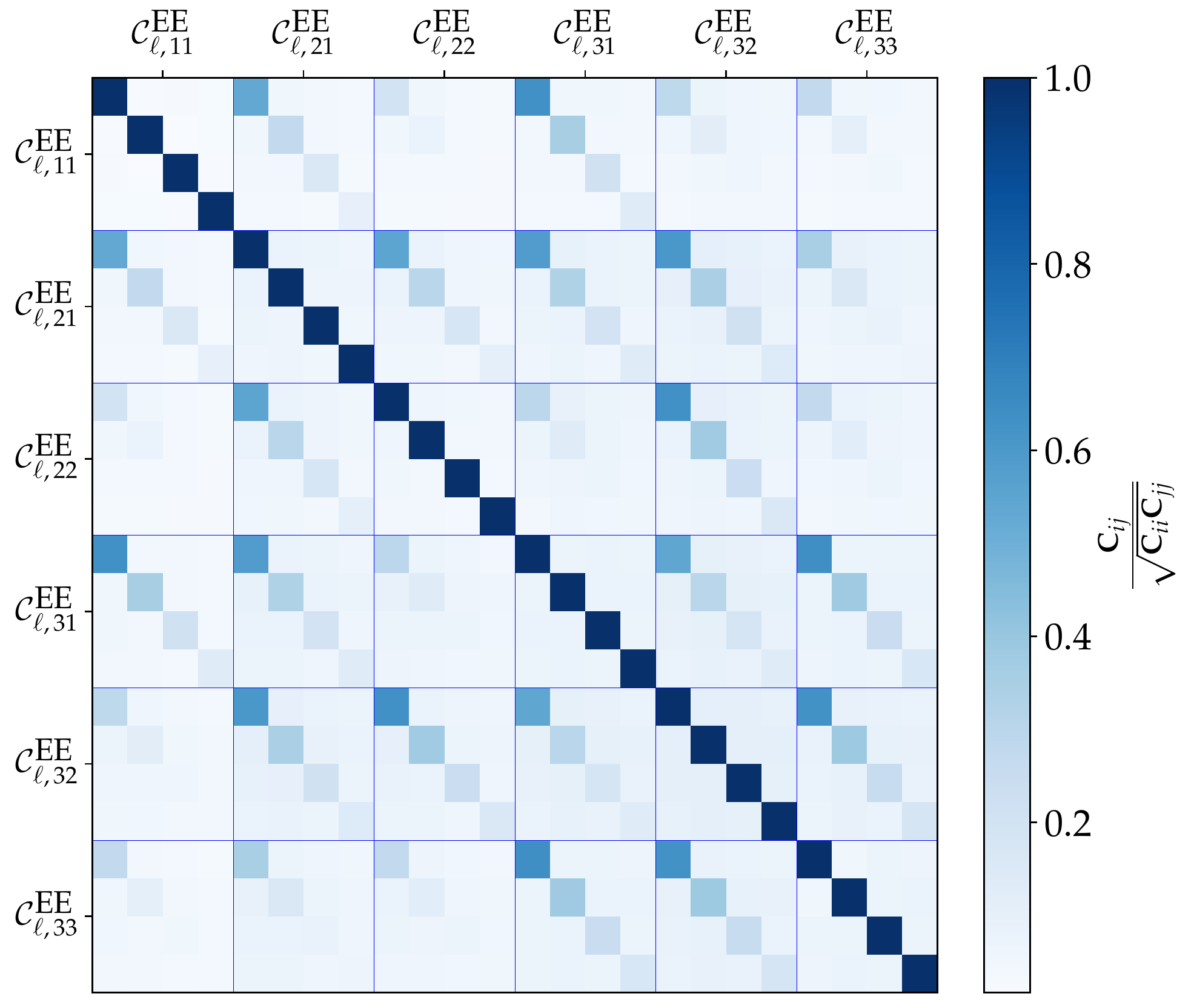}
\par\end{centering}
\caption{\label{fig:kids-covariance}The data correlation matrix for the KiDS-450 analysis. We have ordered the covariance matrix in order of the tomographic labelling $ij$. Note that we have 4 band powers per tomographic bin, hence $6 \times 4$ blocks in the covariance matrix.}   
\end{figure}

\begin{equation}
\label{eq:ard}
k\left(\bs{\theta}_{p},\,\bs{\theta}_{q}\right)= A^{2}\,\tm{exp}\left[-\dfrac{1}{2}\left(\bs{\theta}_{p}-\bs{\theta}_{q}\right)^\tm{T}\bs{\Lambda}^{-1}\left(\bs{\theta}_{p}-\bs{\theta}_{q}\right)\right],
\end{equation}

\noindent where $\bs{\Lambda} = \tm{diag}\left(\lambda_{1},\,\lambda_{2}\ldots\lambda_{d}\right)$. $A$ and $\lambda$ are referred to as the amplitude and characteristic squared length-scale respectively. In particular, the former determines the average distance of the function from the mean while $\lambda$ controls the smoothness of the function. The set of hyper-parameters for this particular kernel is $\{A,\,\bs{\lambda}\}$. This kernel has the nice property that it is fully differentiable and positive definite.

Unlike parametric regression where we define priors over parameters, we will now define a prior covariance over the functions directly and using Bayes' theorem, the joint posterior distribution of the functions is

\begin{equation}
p\left(\bs{f}\left|\right.\bs{\theta},\bs{y}\right)=\dfrac{p\left(\bs{y}\left|\right.\bs{\theta},\bs{f}\right)p\left(\bs{f}\left|\right.\bs{\theta}\right)}{p\left(\bs{y}\left|\right.\bs{\theta}\right)},
\end{equation}

\noindent where the likelihood $p\left(\bs{y}\left|\right.\bs{\theta},\bs{f}\right)\sim\mc{N}\left(\bs{f},\,\bs{\Sigma}\right)$ and the prior is $p\left(\bs{f}\left|\right.\bs{\theta}\right)\sim\mc{N}\left(\bs{0},\sans{K}\right)$. $\bs{\Sigma}$ is the noise covariance covariance matrix. $\sans{K}=k\left(\bs{\theta},\bs{\theta}'\right)$ is the kernel matrix by calculating equation \eqref{eq:ard} for every pair of points in $\bs{\theta}$. Prior to making predictive inference, the Gaussian Process is trained by finding the set of hyper-parameters $\{A,\,\bs{\lambda}\}$ which maximizes the Bayesian Evidence (marginal likelihood)

\begin{equation}
\label{eq:gp_marginal_likelihood}
\tm{log }p\left(\bs{y}\left|\right.\bs{\theta}\right) = -\dfrac{1}{2}\bs{y}^{\tm{T}}\sans{K}_{y}^{-1}\bs{y} - \dfrac{1}{2}\tm{log}\left|\sans{K}_{y}\right| + \tm{constant},
\end{equation}

\noindent where $\sans{K}_{y}=\sans{K}+\bs{\Sigma}$. The first term in the marginal likelihood controls the fit to the data while the second term controls the model complexity. For numerical stability, we first compute the Cholesky factor, $\sans{L}$, of $\sans{K}_{y}\equiv\sans{L}\sans{L}^{\tm{T}}$, solve for $\bs{u}$ in the linear system $\sans{L}\bs{u}=\bs{y}$ followed by solving for $\bs{\alpha}$ in $\sans{L}^{\tm{T}}\bs{\alpha} = \bs{u}$. The marginal likelihood is then given by

\begin{equation}
\label{eq:gp_marginal_likelihood_stable}
\tm{log }p\left(\bs{y}\left|\right.\bs{\theta}\right) = -\dfrac{1}{2}\bs{y}^{\tm{T}}\bs{\alpha} - \sum_{i}\tm{log}\,\sans{L}_{ii} + \tm{constant}.
\end{equation}

Moreover, the partial derivatives of equation \eqref{eq:gp_marginal_likelihood} with respect to the kernel hyperparameters, $\bs{\eta}=\{A,\,\bs{\lambda}\}$ can be computed in closed form

\begin{equation}
\label{eq:grad_kernel}
\dfrac{\partial}{\partial\bs{\eta}_{i}}\tm{log }p\left(\bs{y}\left|\right.\bs{\theta}\right) = \dfrac{1}{2}\tm{tr}\left[\left(\bs{\alpha}\bs{\alpha}^{\tm{T}}-\sans{K}^{-1}\right)\dfrac{\partial\sans{K}}{\partial\bs{\eta}_i}\right]
\end{equation}

\noindent and $\bs{\alpha}=\sans{K}_{y}^{-1}\bs{y}$. The gradients are useful when maximising the marginal likelihood when using gradient-based optimisation. Another option, which we will not use for this particular application, is to marginalise over the kernel hyperparameters (given an appropriate set of priors) in a fully Bayesian formalism. The reason for not taking this route is that the number of latent variables in the data model becomes large $(\sim 10^{2})$ and recall that the marginal likelihood for a GP is an expensive calculation.

For a given test point $\bs{\theta}_{*}$, the posterior distribution of the function, $f_{*}=f\left(\bs{\theta}_{*}\right)$ is a Gaussian distribution with mean and variance given by

\begin{equation}
\label{eq:predictionGP}
\begin{split}
\bar{f_{*}}&= \bs{k}_{*}^{\tm{T}}\bs{\alpha} % \bs{k}_{*}^{\tm{T}}\sans{K}_{y}^{-1}\bs{y} =
\\
\tm{var}\left(f_{*}\right)&=\bs{k}_{**}-\bs{k}_{*}^{\tm{T}}\sans{K}_{y}^{-1}\bs{k}_{*}.
\end{split}
\end{equation}

The GP approach is quite appealing as it predicts both the mean and variance, at the expense of an $\mc{O}(N^{2})$ operation, of the function for a particular test point. Moreover, as seen from equation \eqref{eq:predictionGP}, the mean of a GP is a linear predictor. Once $\bs{\alpha}$ is calculated, the mean function can be quickly and easily calculated for any set of test points since it involves only $\mc{O}(N)$ operations. In addition, if required, the analytical gradient with respect to the mean function at a particular test point can also be calculated. However, a GP can be prohibitively expensive for large data sets since it involves the computation of the Cholesky factor which has computational complexity of $\mc{O}(N^{3})$ during the training phase and requires $\mc{O}(N^2)$ operations for the predictive variance. Besides, a GP has $\mc{O}(N^{2})$ memory requirements for storing the Cholesky factor (if the computation of the predictive error is required) and the vector $\alpha$. 

In the next section, we look into how a GP emulator can be useful for future weak lensing surveys, for which we naively expect around 10 tomographic redshift bins and thousands of summary statistics \citep{EuclidFisher}. In particular, one can either emulate the band powers directly or the MOPED coefficients. We investigate both and discuss the advantages of using the MOPED coefficients in the following sections.

\begin{table}
\caption{\label{table:inputs}Set of cosmological and systematic parameters which are used as inputs in the emulator. $\bs{\theta}$ is the set of parameters (first 8 parameters in the table below) used for the emulation scheme and $\bs{\beta}$ is the set of the remaining 4 parameters which are also marginalised over.}
\renewcommand*{\arraystretch}{1.5}
\begin{tabular*}{0.45\textwidth}{lc}
\hline 
Definition & Symbol\tabularnewline
\hline 
CDM density & $\Omega_{\tm{cdm}}h^{2}$\tabularnewline
Baryon density & $\Omega_{\tm{b}}h^{2}$\tabularnewline
Scalar spectrum amplitude & $\tm{ln}(10^{10}A_{\tm{s}})$\tabularnewline
Scalar spectral index & $n_{s}$\tabularnewline
Hubble parameter & $h$\tabularnewline
Free amplitude baryon feedback parameter & $A_{\tm{bary}}$\tabularnewline
Intrinsic alignment parameter & $A_{\tm{IA}}$\tabularnewline
Neutrino mass (eV) & $\Sigma m_{\nu}$\tabularnewline
\hline 
Free amplitude (bin 1) & $A_{1}$ \tabularnewline
Free amplitude (bin 2) & $A_{2}$  \tabularnewline
Free amplitude (bin 3) & $A_{3}$\tabularnewline
Multiplicative bias & $m$ \tabularnewline
\hline 
\end{tabular*}
\end{table}
\section{Emulator}
\label{sec:emulator}
In this section, we use the formalism presented above to build the emulator. In brief, the latter involves 4 main stages, 1) generating a set of design points, 2) running the full forward simulator at these points, 3) training the emulator and 4) making predictions at test locations in the parameter space. Once this is done, the emulator is connected to an MCMC sampler to obtain the marginalised posterior distributions of the parameters in our model. A simple flow of the core idea is shown in Fig. \ref{fig:idea-pipeline}. In the following, we touch briefly on the data we have used for our analysis before systematically going through the steps we have taken to build the emulator. 

\subsection{Data}

We use the publicly-available weak lensing data from \cite{2017MNRAS.471.4412K} to test the performance of our emulator. We use 3 tomographic redshift bins, namely, $0.10<z<0.30$, $0.30<z<0.60$ and $0.60<z<0.90$ and the convergence power spectrum is computed in the range $10<\ell<4000$. Moreover, we follow \cite{2017MNRAS.471.4412K} and drop the first, second-to-last and last band powers in our analysis, that is, we use only the band powers corresponding to the following $\ell$-ranges:$76\leq\ell<220$, $221\leq\ell<420$, $421\leq\ell<670$ and $671\leq\ell<1310$. For a 3-bin tomographic analysis, we have 6 auto- and cross- tomographic power spectra to calculate. The data and covariance matrix for this problem are shown in Fig. \ref{fig:kids-data} and \ref{fig:kids-covariance} respectively. 

The emulator can be built at the level of the power spectra or the band powers. Here we choose to build a GP for each band power, giving 24 GPs. Alternatively, for likelihood-free inference methods, one can also emulate the likelihood directly using the GPs (see \cite{2018PhRvD..98f3511L} and \cite{2007arXiv0712.0194F}). For power spectrum reconstruction, one can use the PICO method or an alternative, but constrictive, stance is to adopt the approach taken by \cite{2007PhRvD..76h3503H} to first learn a set of basis functions via Singular Value Decomposition (SVD) and model the resulting weights by a Gaussian Process. However, building an emulator for weak lensing analysis needs to account for systematic effects, but some of these can be included analytically without emulation, resulting in an 8-dimensional GP, rather than 12 (6 cosmological and 6 systematic parameters) if we were to emulate the likelihood.

\begin{figure}
\noindent \begin{centering}
\includegraphics[width=0.45\textwidth]{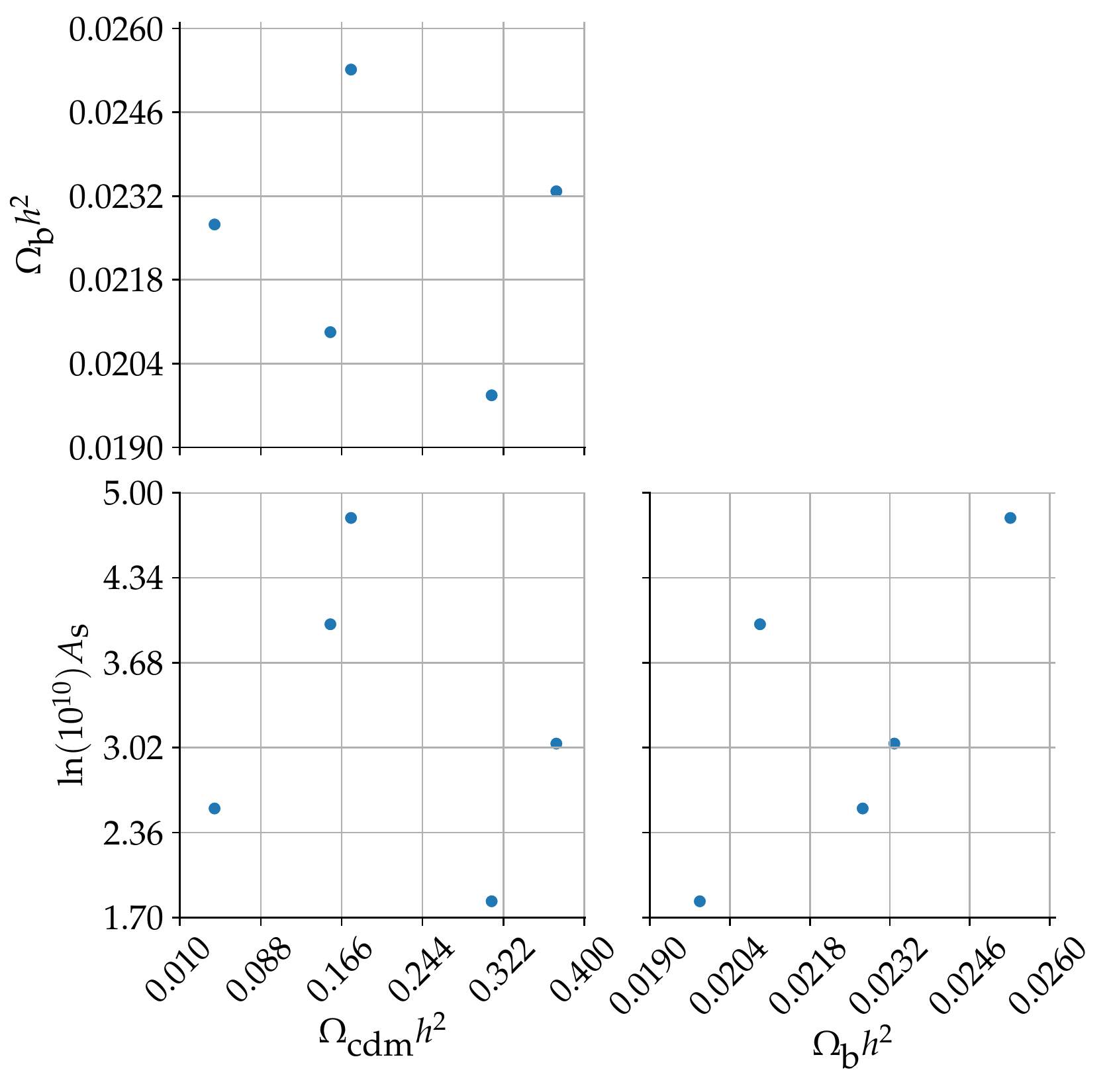}
\par\end{centering}
\caption{\label{fig:lhs_corner}Five Latin Hypercube samples (using the maximin method) projected in 2D. In particular, we generate five Latin Hypercube samples in 8D and we scale them according to our pre-defined priors. In the figure, we show the projection in 2D for 3 parameters and as expected, each point occupies it corresponding row and column.}
\end{figure}

\begin{figure}
\noindent \begin{centering}
\includegraphics[width=0.4\textwidth]{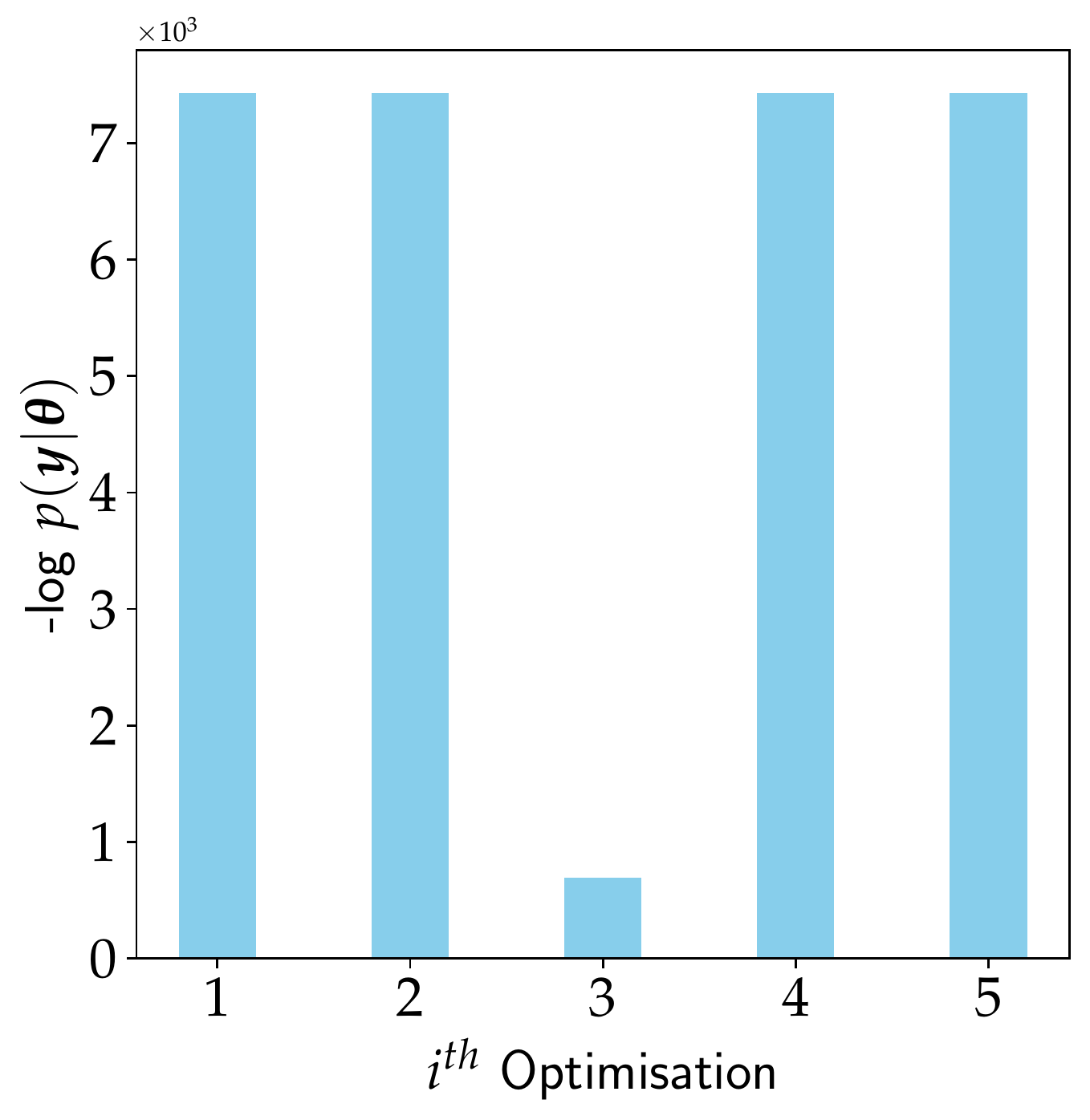}
\par\end{centering}
\caption{\label{fig:optimisation}The figure shows the marginal likelihood of the Gaussian Process (with 3000 training points) for the fourth band power matrix and $i=j=2$. Note the local minimum for the $3^{\tm{rd}}$ run of the optimiser. The other bars have almost the same value, hence showing that $N_{\tm{restart}}=5$ is a good choice for training the GP.}
\end{figure}

\subsection{Training Points}
The generation of the training points is a key ingredient for the emulator to perform well. Accurate high-dimensional regression is not easy, mainly due to the curse of dimensionality. With the formalism presented in this work and depending on the complexity of the function, one can reconstruct the function precisely and accurately in low dimensions, hence leading to an accurate likelihood as would be the case if we were to use the full simulator, CLASS \citep{2011arXiv1104.2932L} in this case. As the dimensionality of the problem increases, we need an exponentially increasing number of training points to emulate the true function accurately.  

In PICO, the training points were generated uniformly from a box whose sides were centred on the mean of a converged MCMC chain (consisting of $\sim 60000$ cosmological models) and width $3\sigma$ along each direction. In the second release of PICO, they selected training points which lie within 25 log-likelihoods of the WMAP peak \citep{2007arXiv0712.0194F}. On the other hand, \cite{2007MNRAS.376L..11A} first drew 2000 training points from the same box defined in PICO and also added an extra 5000 training points drawn from a Gaussian distribution, whose covariance was twice the expected covariance matrix, centred on the maximum likelihood. These techniques perform quite well for two reasons: 1) by restricting the prior volume of the training points to the high likelihood regions allows the sampler explictly to explore this specific region in parameter space, 2) creating a data set with thousands of training points will also improve \textit{any} regression method. A shortcoming of using these approaches is that the algorithm will not perform well in regions where there is no training point nearby (see Appendix A in \cite{2007PhRvD..76h3503H} for a comparison of their method with PICO). This is a typical manifestation of almost any Machine Learning algorithm. They are good at making reliable predictions within a pre-defined prior, provided they are trained with enough data points. Building Machine Learning algorithms in the small data regime is still in its infancy, hence an active area of research \citep{2019arXiv190109054B}. 

Moreover, if the training points are naively generated randomly from our pre-defined priors, we might not obtain a suitable coverage of the parameter space. A possible solution to this, is to use a grid but then the number of training points grows exponentially as the dimensionality of the problem increases. As an example, say, we have a 7D problem and we choose to have 10 points per parameter, then our training set will have 10 million points. 

Alternatively, we can use Latin Hypercube (LH) sampling \citep{mckay1979comparison} which is a method for generating random samples from a multidimensional dimensional distribution in a controlled (quasi-random) way. A point is assigned such that it uniquely occupies its row and column respectively. This procedure generalises to higher dimensional designs. In Fig. \ref{fig:lhs_corner}, we show the projection of 5 LH samples, which have been generated from a box in 8D and scaled by the pre-defined priors in \S\ref{sec:priors}, in 2D. In particular, we show the projection for 3 parameters only but the same applies for the other parameters, where each point uniquely occupies its corresponding row and its column. The LH method is now a ubiquitous tool for performing emulation in large simulation scenarios \citep{2007PhRvD..76h3503H, 2011ApJ...728..137S, 2018MNRAS.475.1213S} and is seen to be quite efficient, not only in producing a fair interpolation, but also provides reasonable posterior densities. 

In this work, we adopt the LH approach to generate our training set. The LH samples are generated using the \texttt{maximinLHS} function from the \texttt{lhs} R package \citep{carnell2012lhs}. This particular design relies on distance criterion \citep{JOHNSON1990131} and the final design is a result of maximising the minimum distance between points.

\subsection{Priors}
\label{sec:priors}
In our baseline emulator, we generate 1000 Latin Hypercube samples from a box, between 0 and 1. We first linearly transform these samples to the range of the pre-defined prior box for the 6 cosmological and 2 systematics parameters,

\begin{equation*}
\begin{split}
%\Omega_{b}h^{2},\,ln(10^{10}A_{s}),\,n_{s},\,h,\,A_{bary},\,\sum\nu
\Omega_{\tm{cdm}}h^{2}&\sim\mc{U}[0.01, 0.40]\\
\Omega_{b}h^{2}&\sim\mc{U}[0.019, 0.026]\\
\tm{ln}(10^{10}A_{s})&\sim\mc{U}[1.70, 5.00]\\
n_{s}&\sim\mc{U}[0.70, 1.30]\\
h&\sim\mc{U}[0.64, 0.82]\\
A_{\tm{bary}}&\sim\mc{U}[0.0, 2.0]\\
A_{\tm{IA}}&\sim\mc{U}[-6.0, 6.0]\\
\Sigma m_{\nu}&\sim\mc{U}[0.06, 1.0]
\end{split}
\end{equation*}

\noindent followed by running the full simulator at these points to obtain the total band powers. $\mc{U}[a,\,b]$ denotes a uniform distribution with lower and upper limits $a$ and $b$ respectively. We apply a more restrictive prior than the original KiDS-450 prior [0.01,0.99] for $\Omega_{\tm{cdm}}h^{2}$ since otherwise a large fraction of the LH samples we generate lie outside the region of parameter space constrained by the current weak lensing analysis. Moreover, having a smaller volume of parameter space also improves the performance of the emulator. The prior for the $A_{\tm{bary}}$ is set to an upper limit of 2 (instead of 10 in \cite{2017MNRAS.471.4412K}) because we found that, values of $A_{\tm{bary}}\gtrsim 3$ lead to negative $b^{2}$, which implies an unphysical negative power spectrum. In the same spirit, large values of $A_{\tm{bary}}$ lead to negative auto-correlated band powers and in some cases, the band power matrix (equation \eqref{eq:band_power_matrix}) was not positive definite. We also found that large values of neutrino masses, $\Sigma m_{\nu} \gtrsim 1 \tm{eV}$ result in almost half of the CLASS band powers in our training set to be \texttt{nan}. We therefore set an upper limit for $\Sigma m_{\nu}$ to 1 eV.

\subsection{Transformations}

Training the Gaussian Processes with the LH samples from above might be suboptimal, the reason being that the volume occupied by a hypercube grows exponentially with increasing dimensions. On the other hand, a sphered training set (hypersphere) has a smaller volume compared to its corresponding hypercube but with the same scaling with dimension. This transformation step is analogous to the one used by \cite{2007ApJ...654....2F}. \cite{2011ApJ...728..137S} assessed in detail the effect of various transformations prior to building an emulator for the CMB power spectrum. They found that de-correlating the input space leads to significant improvements compared to working with the original form of the input parameters. The interpolation can further be improved if one uses a known Fisher information matrix specific to the problem. 

The transformation matrix can be calculated as follows: we first compute the sample covariance, $\sans{C}_{\bs{\theta}}$ of the 1000 input parameters, $\bs{\theta}$ to the emulator (see Table \ref{table:inputs}), which we diagonalise, $\sans{C}_{\bs{\theta}} = \sans{U}\sans{D}\sans{U}^{\tm{T}}$.  $\sans{U}$ is a $d\times d$ orthonormal matrix and $\sans{D}$ is a diagonal $d\times d$ matrix consisting of the (necessarily positive) eigenvalues. The transformation matrix which whitens $\bs{\theta}$ is then $\sans{U}\sans{D}^{\frac{1}{2}}$, such that the transformed input covariates are
%\begin{equation}
$\sans{X} = \sans{U}\sans{D}^{\frac{1}{2}}\bs{\theta}$,
%\bs{\theta}_{\tm{sphered}}
%\end{equation}
and the covariance of $\sans{X}$ is the identity matrix. Also, having a pre-whitened basis also justifies the use of a diagonal kernel matrix such as the ARD kernel in equation \eqref{eq:ard}, for which it is often blindly assumed (without transforming the inputs) that the correlation among the input parameters vanishes.

Next, we consider the transformation of the band powers. The distribution of the original band powers in our training set is left-skewed. For a fixed $\ell$ in our 3-bin tomographic analysis, the resulting $3\times 3$ matrix, 

\begin{equation}
\label{eq:band_power_matrix}
\sans{B}_{\ell}=\left(\begin{array}{ccc}
\tm{B}_{\ell,\,00} & \tm{B}_{\ell,\,01} & \tm{B}_{\ell,\,02}\\
\tm{B}_{\ell,\,10} & \tm{B}_{\ell,\,11} & \tm{B}_{\ell,\,12}\\
\tm{B}_{\ell,\,20} & \tm{B}_{\ell,\,21} & \tm{B}_{\ell,\,22}
\end{array}\right)
\end{equation}

\noindent must be positive-definite and emulating the matrix elements individually will not guarantee this.

\begin{figure*}
\noindent \begin{centering}
\includegraphics[width=0.9\textwidth]{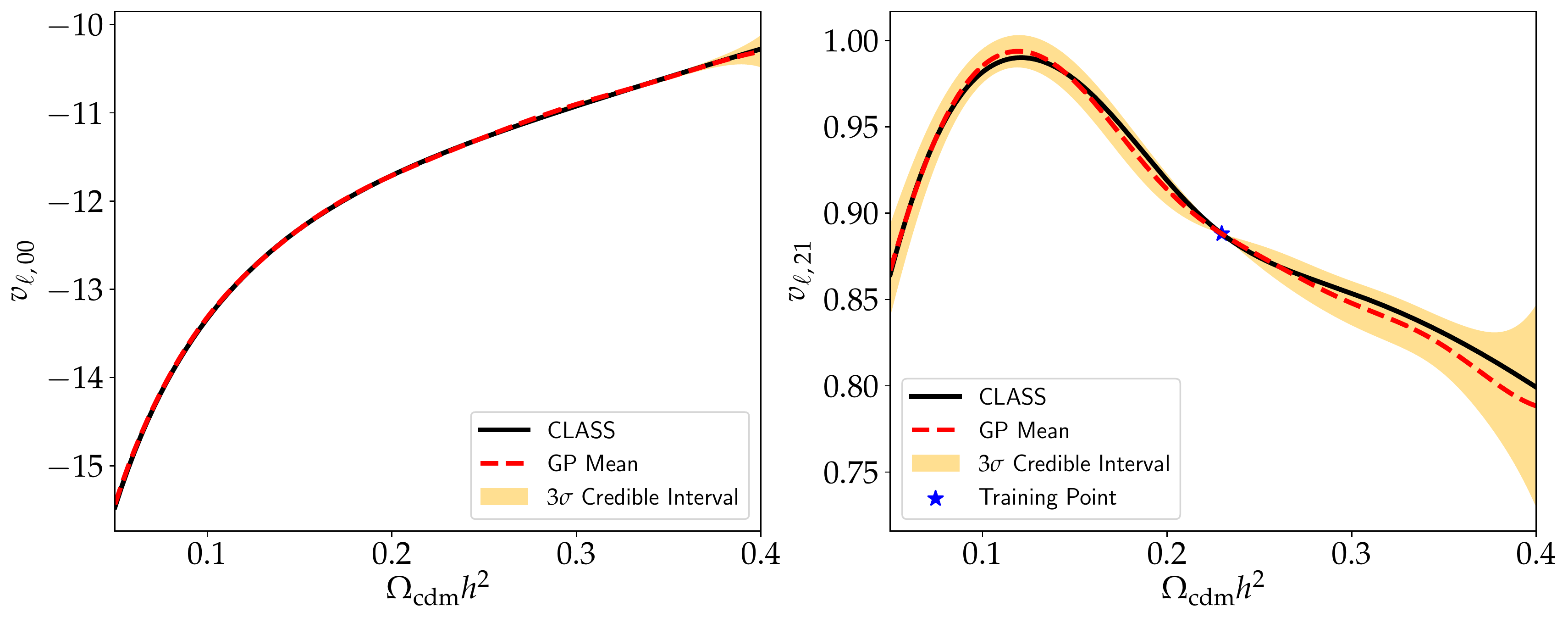}
\par\end{centering}
\caption{\label{fig:slice}The left plot shows the predicted band power across a slice in parameter space. In other words, we choose a point within the prior box and compute the GP mean, variance and the actual band power for $\Omega_{\tm{cdm}}h^{2}\in[0.05, 0.40]$. The same procedure is repeated in the right plot, but we instead choose a point from a training set, to illustrate the fact that the predicted GP uncertainty tends to zero near the training point and the predictive variance increases towards the edge of the prior box.}
\end{figure*}

To ensure that the $3\times 3$ band power matrix remains positive-definite during the prediction phase when using the emulator, we instead build the latter on each element of the logarithm $\sans{B}_{\ell}$ (lower or upper triangular part, essentially all the unique elements),  

\begin{equation}
\label{eq:target_functions}
\sans{V}_{\ell} = \sans{R}\tilde{\bs{\Lambda}}\sans{R}^{\tm{T}} = \tm{log}\left(\sans{B}_{\ell}\right),
\end{equation}

\noindent where $\sans{B}_{\ell}=\sans{R}\bs{\Lambda}\sans{R}^{\tm{T}}$,  $\tilde{\bs{\Lambda}}_{\nu\nu}=\tm{log}(\bs{\Lambda}_{\nu\nu})$ and $\bs{\Lambda}$ and $\tilde{\bs{\Lambda}}$ are diagonal. Moreover, since we normally assume a Gaussian Process with mean zero and kernel, $\sans{K}$, we do an additional linear scaling such that the mean of the band powers in our training set is zero and has a standard deviation of one, for example, for the $i^{th}$ transformed band power,

\begin{equation}
\bs{v}'_{i}\rightarrow\dfrac{\bs{v}_{i} - \bar{v}_{i}}{\sigma_{i}}
\end{equation}

\noindent and the predictive mean and variance are
\begin{equation}
\label{eq:prediction_scaled}
\begin{split}
\bb{E}[v_{i(*)}]   &= \sigma_{i}\,\bb{E}[v'_{i(*)}] + \bar{v}_{i}\\
\tm{var}[v_{i(*)}] &= \sigma_{i}^{2}\,\tm{var}[v'_{i(*)}].
\end{split}
\end{equation}

\subsection{Training the Emulator}

We now have our training set $\{\sans{X},\sans{V}_{\ell,ij}\}$. Therefore we have a set of 24 Gaussian Processes due to each element of the transformed band powers. Prior to building the emulator, a crucial step is to choose a kernel function for the Gaussian Process. Here we use the ARD kernel, defined in equation \eqref{eq:ard}.  

To ensure a good performance, we have to find the set of hyperparameters which maximises the marginal likelihood, as discussed in \S\ref{sec:gp}. An important ingredient is the analytical gradient of the marginal likelihood with respect to the kernel hyperparameters to guarantee convergence to the global minimum. The gradients are

\begin{equation}
\begin{split}
\dfrac{\partial k_{pq}}{\partial A} &= \dfrac{2}{A}k_{pq}^{\tm{ARD}}\\
\dfrac{\partial k_{pq}}{\partial \ell_{i}} &= k_{pq}^{\tm{ARD}}\dfrac{(\bs{\theta}_{p(i)} - \bs{\theta}_{q(i)})^{2}}{\ell_{i}^{3}},
\end{split}
\end{equation}

\noindent where $i$ indicates the $i^{th}$ dimension of the problem. We use the  Limited memory Broyden-Fletcher-Goldfarb-Shanno, L-BFGS-B algorithm \citep{10.1145/279232.279236, press2007numerical} along with the gradients defined above to optimise for these hyperparameters by minimising the negative log-marginal likelihood, in equation \eqref{eq:gp_marginal_likelihood}, via gradient descent. However, it is a known fact that training a Gaussian Process is not an easy task because the marginal likelihood has various local maxima \citep{2006gpml.book.....R}. We adopt the standard approach of restarting our optimiser at different positions and we find that $N_{\tm{restart}}=5$ was sufficient in practice to ensure that we find the set of hyperparameters corresponding to the global optimum (see Fig. \ref{fig:optimisation}). Although this is not guaranteed, we also want to emphasise that the use of the gradients was required to find the global optimum. Once the Gaussian Process is trained, the kernel parameters are fixed at the optimised values of the hyperparameters and then use equations \eqref{eq:predictionGP} to make predictions.

\subsection{The GP Uncertainty}
In this section, we look into propagating the GP uncertainty through the full forward model when we use the emulator. To be more specific, we seek the posterior distributions of the cosmological parameters and the two nuisance parameters $(A_{\tm{IA}},\,A_{\tm{bary}})$, that is, 

$$\bs{\theta}=\left[\Omega_{\tm{cdm}}h^{2},\,\Omega_{\tm{b}}h^{2},\,\tm{ln}(10^{10}A_{\tm{s}}),\,n_{s},\,h,\,A_{\tm{bary}},\,A_{\tm{IA}},\,\Sigma m_{\nu}\right]$$

\noindent and the other 4 nuisance parameters, 

$$\bs{\beta}=\left[A_{1},\,A_{2},\,A_{3},\,m\right]$$

\noindent marginalised over the probabilistic band powers. $A_{1},\,A_{2},\,A_{3}$ correspond to free parameters which determine excess noise in the autocorrelation power spectrum, while $m$ is the shear multiplicative bias parameter \citep{2017MNRAS.471.4412K}. Using equation \eqref{eq:cl_tot} and defining $\bs{v}$ as the total band powers, we can write the joint posterior, $p(\bs{\theta},\,\bs{\beta}\left|\bs{d}\right.)$ as

\begin{equation}
\begin{split}
p\left(\bs{\theta},\,\bs{\beta}\left|\bs{d}\right.\right) &= \int p\left(\bs{\theta},\,\bs{\beta},\,\bs{v}\left|\bs{d}\right.\right)\,\tm{d}\bs{v}\\
&=\int p\left(\bs{d}\left|\bs{v},\,\bs{\beta}\right.\right)p\left(\bs{v}\left|\bs{\theta}\right.\right)\,\tm{d}\bs{v}\,p\left(\bs{\theta}\right)p\left(\bs{\beta}\right).
\end{split}
\end{equation}

\noindent If $p(\bs{\nu}|\bs{\theta})$ were a Gaussian distribution of the band power from the Gaussian Process, the above integration would be a convolution of two Gaussian distributions and the likelihood part would be Gaussian.

\begin{figure}
\noindent \begin{centering}
\includegraphics[width=0.20\textwidth]{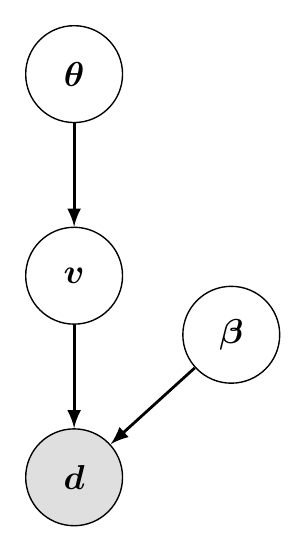}
\par\end{centering}
\caption{\label{fig:dag}The full forward model can be understood as follows: at each step in the inference procedure, a random set of samples of the cosmological, $\bs{\theta}$ and nuisance, $\bs{\beta}$ is drawn from the prior, followed by a random realisation of the probabilistic band powers, centred on its mean and variance before computing the likelihood. Note that the kernel hyperparameters are fixed to their optimised values.}   
\end{figure}

\begin{figure*}
\noindent \begin{centering}
\includegraphics[width=0.9\textwidth]{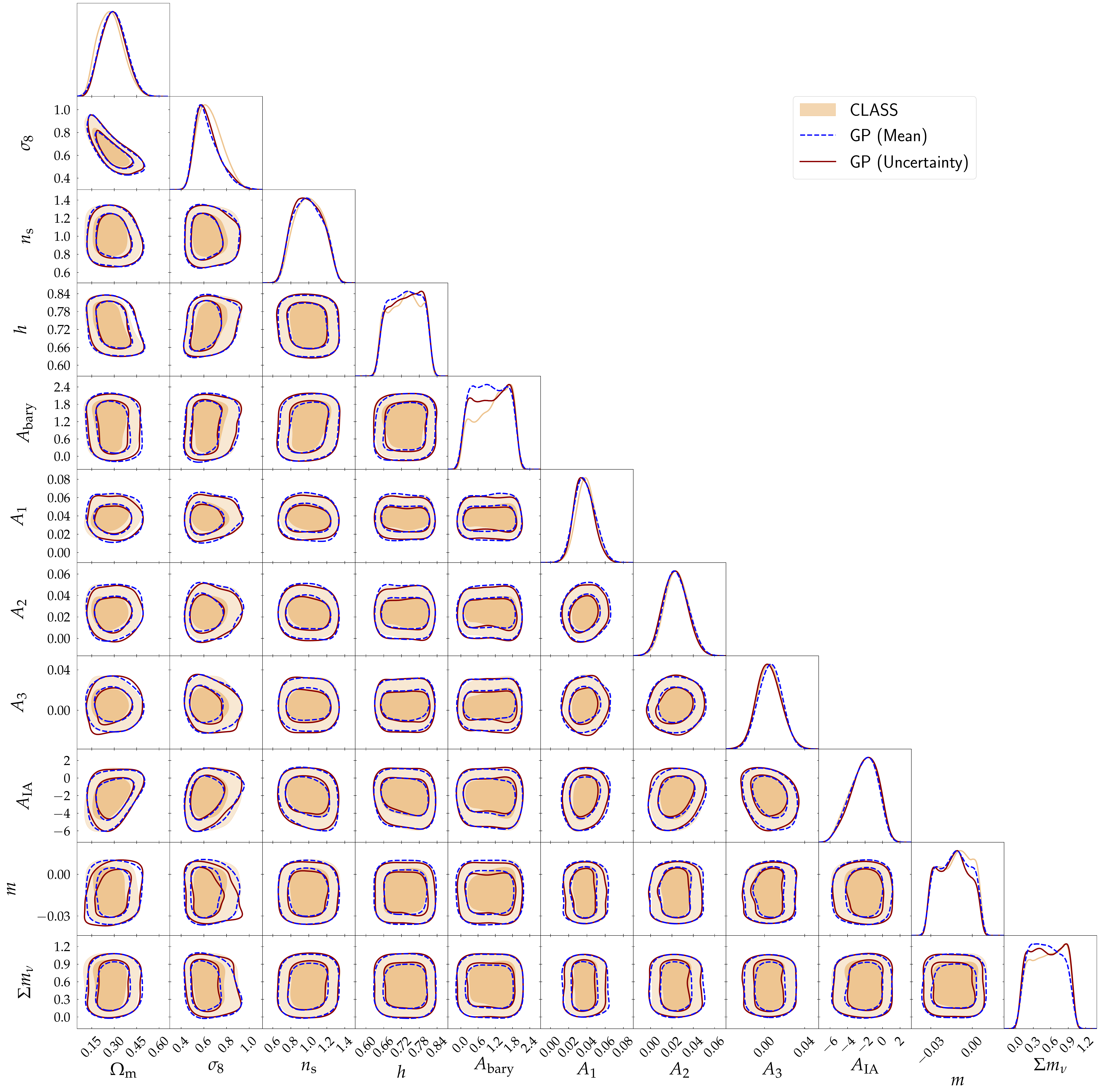}
\par\end{centering}
\caption{\label{fig:posterior_plot}The full 1D and 2D marginalised posterior distributions obtained using three different methods - The one in tan colour corresponds to posterior distributions with the full simulator (CLASS) while the solid brown one corresponds to the Gaussian Process emulator when random functions of the band powers are drawn, hence marginalising over the Gaussian Process uncertainty. The posterior in blue shows the distributions obtained when only the mean of the Gaussian Process was used in the inference routine. The contours denote the 68 \% and 95 \% credible interval respectively. Note that some parameters are dominated by their respective priors and are not constrained at all. A similar conclusion was drawn by \citep{2017MNRAS.471.4412K}. However, the important point here is that the posterior from the GP is close to that obtained with CLASS.}
\end{figure*}

However, in our analysis, the predictive distribution is Gaussian in each element of the logarithm of the band power matrix. For example, in Fig. \ref{fig:slice}, we show the GP mean and variance for two elements across a slice in parameter space. As previously discussed, if the GP predictions were Gaussian in the band powers, we could marginalise analytically over the theoretical uncertainty. Since they are Gaussian in each element of $\sans{V}_{\ell,ij}$, we marginalise by drawing samples of the cosmological and nuisance parameters (see Fig. \ref{fig:dag}) and perform a Monte Carlo integration, which is relatively fast and approximating the joint posterior as

\begin{equation}
\label{eq:monte_carlo_integral}
p(\bs{\theta},\,\bs{\beta}\left|\bs{d}\right.)\propto\dfrac{p(\bs{\theta})p(\bs{\beta})}{N_{s}}\sum_{i=1}^{N_{s}}p\left(\bs{d}\left|\sans{V}_{\ell,ij}\right.\right).
\end{equation}

\noindent $N_{s}$ is the number of random band powers drawn after computing the predictive mean and variance. We use $N_{s}=20$ at every step in the MCMC to take into account the uncertainty from the Gaussian Process. Recall that each band power is being modelled independently by a GP and hence the Monte Carlo integral in equation \eqref{eq:monte_carlo_integral} requires few draws of the probabilistic band powers. 
\section{Data Compression}
\label{sec:data_compression}

\begin{figure*}
    \centering
    \subfloat[log-Posterior without Compression]{{\includegraphics[width=0.45\textwidth]{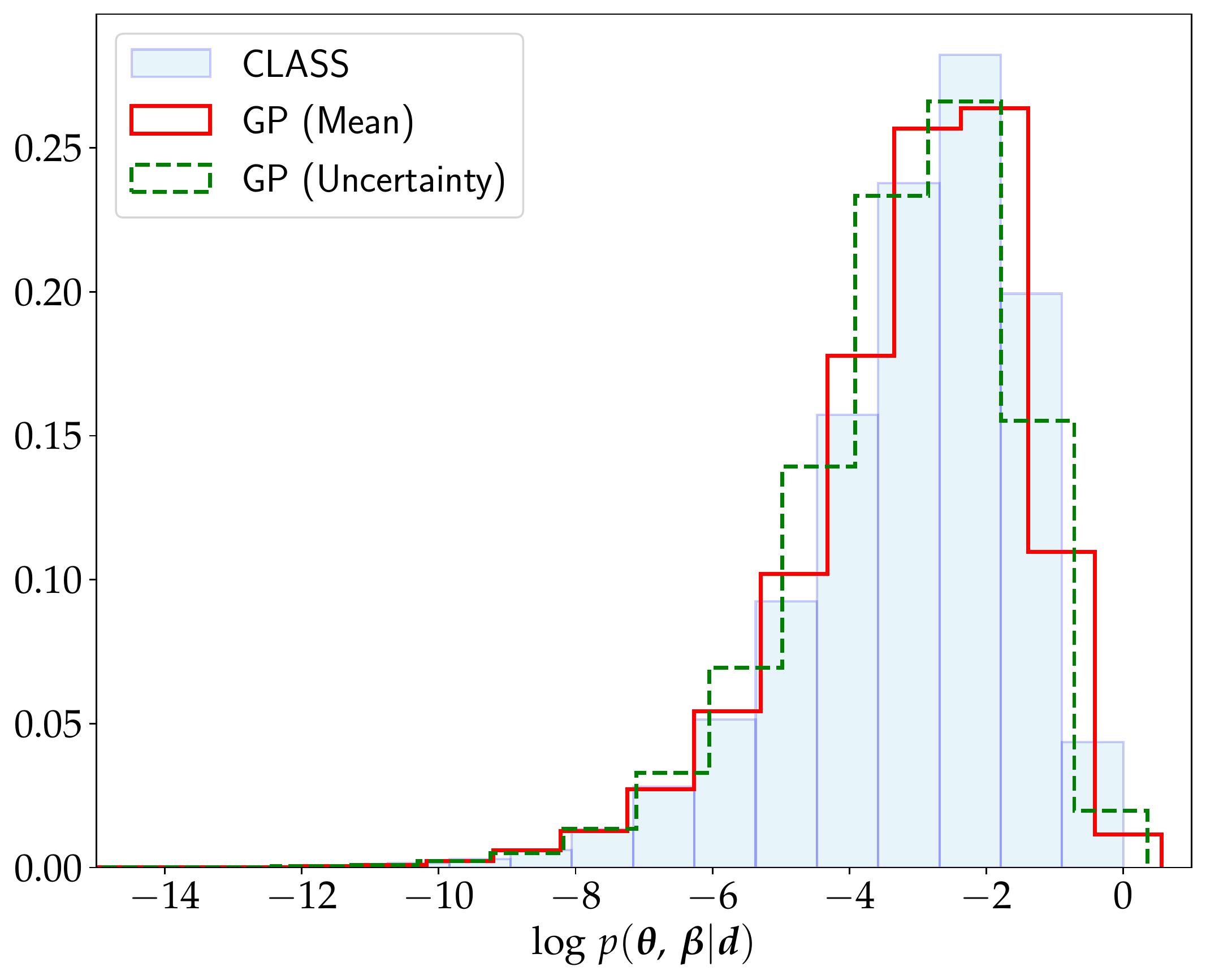}}}
    \qquad
    \subfloat[log-Posterior with MOPED Compression]{{\includegraphics[width=0.45\textwidth]{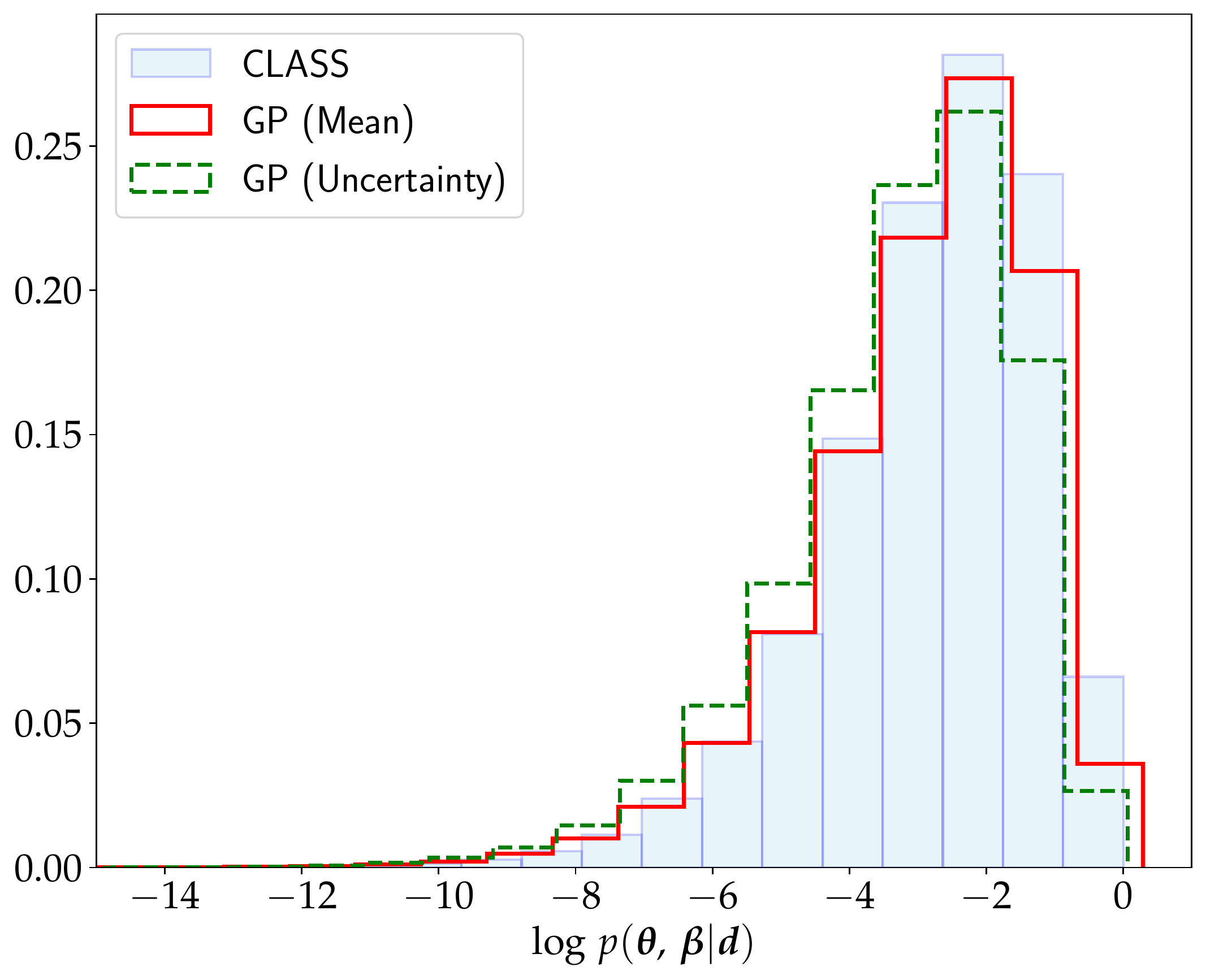}}}
    \caption{\label{fig:post_samples}Samples of the log-posterior obtained with the 3 methods investigated - In panel (a), the pale blue histogram refers to the log-posterior samples from CLASS while the red and green step histogram correspond to the mean and error of the GP respectively. A similar plot is shown in panel (b) but after applying the MOPED compression step.}
\end{figure*}

The next era of weak lensing surveys such as Euclid and LSST will have $\sim 10$ tomographic bins, and with multiple band powers or correlation functions per bin, the number of summary statistics will be large, $\sim 10^{3}-10^{4}$.  As an example, \cite{EuclidFisher} considered 100 bandpowers per bin, and 10 tomographic bins, which gives a minimum of 1000 summaries, and 5500 if cross-band powers are included. The setup, in the previous section, is not a scalable approach for these future surveys. In particular, emulating each band power is not an entirely feasible approach because one will have to train and store thousands of separate Gaussian Processes and this process in itself can be quite expensive.

In this section, we show that the emulator can be used with the MOPED algorithm \citep{2000MNRAS.317..965H} which reduces the number of data points from $N$ to just $p$ numbers. $N$ is the number of data points and $p$ is the number of parameters in our model. For current weak lensing analysis, the gain is not significant (since we are working with only 24 band powers) but the method proposed in this work is expected to yield fast parameter inference in the regime of a large number of band powers, $N\sim10^{4}$, with only $p\sim 10$ parameters of interest.

Here, we briefly cover the MOPED algorithm. The latter essentially finds some weighing vector, $\bs{b}$ which encapsulates as much information as possible for a specific model parameter $\bs{\theta}_{\alpha}$. This vector is then used to find linear combination of the data, $\bs{d}$ such that the compressed data is

\begin{equation}
y_{\alpha}\equiv\bs{b}^{\tm{T}}_{\alpha}\bs{d}.
\end{equation}

\noindent The first and subsequent MOPED vectors are given respectively by

\begin{equation}
\bs{b}_{1}=\dfrac{\sans{C}^{-1}\bs{\mu}_{,1}}{\sqrt{\bs{\mu}_{,1}^{\textrm{T}}\sans{C}^{-1}\bs{\mu}_{,1}}}
\end{equation}

\noindent and 

\begin{equation}
\bs{b}_{\alpha}=\dfrac{\sans{C}^{-1}\bs{\mu}_{,\alpha}-\sum_{\beta=1}^{\alpha-1}(\bs{\mu}_{,\alpha}^{\textrm{T}}\bs{b}_{\beta})\bs{b}_{\beta}}{\sqrt{\bs{\mu}_{,\alpha}^{\textrm{T}}\sans{C}^{-1}\bs{\mu}_{,\alpha}-\sum_{\beta=1}^{\alpha-1}(\bs{\mu}_{,\alpha}^{\textrm{T}}\bs{b}_{\beta})^{2}}}\;\;(\alpha>1),
\end{equation}

\noindent where $\sans{C}$ is the data covariance matrix and $\bs{\mu}_{,\alpha}$ is the vector obtained by calculating the gradient of our theoretical model at a fiducial parameter set. In the previous applications of the MOPED algorithm, it was assumed that the covariance matrix is fixed. In our case, \cite{2017MNRAS.471.4412K} constucted a covariance matrix which depends on the $m$ parameter, the multiplicative bias. In this work, we fix $\sans{C}$ at the average fiducial value provided\footnote{Cosmological parameter inference depends mildly on the parameter $m$.} in the data. Data compression with parameter-dependent covariance matrix has been explored by \cite{2017MNRAS.472.4244H}. If $\sans{B}\in\bb{R}^{N\times p}$ is the matrix whose columns consist of the MOPED vectors, the compressed data vector is just

\begin{equation}
\bs{y}=\sans{B}^{\tm{T}}\bs{x}.
\end{equation}

\noindent By construction, the MOPED vectors $\bs{b}_{\alpha}$ are orthogonal to each other, that is, $\bs{b}_{\alpha}^{\tm{T}}\sans{C}\bs{b}_{\beta}=\delta_{\alpha\beta}$. Therefore, the covariance matrix of $\bs{y}$, $\sans{B}^{\tm{T}}\sans{C}\sans{B}=\bb{I}$, the identity matrix, of size $p\times p$. As a result of this orthogonality condition, elements from the compressed data vector are uncorrelated. Hence, the log-likelihood is straightforwardly

\begin{equation}
\label{eq:moped_likelihood}
\tm{log}\,\mc{L} = -\dfrac{1}{2}\sum_{\alpha=1}^{p}(y_{\alpha}-\bs{b}_{\alpha}^{\tm{T}}\bs{\mu})^{2}  + \tm{constant},
\end{equation}

\begin{figure*}
    \centering
    \subfloat[Emulator for computing $\sigma_{8}$]{{\includegraphics[width=0.30\textwidth]{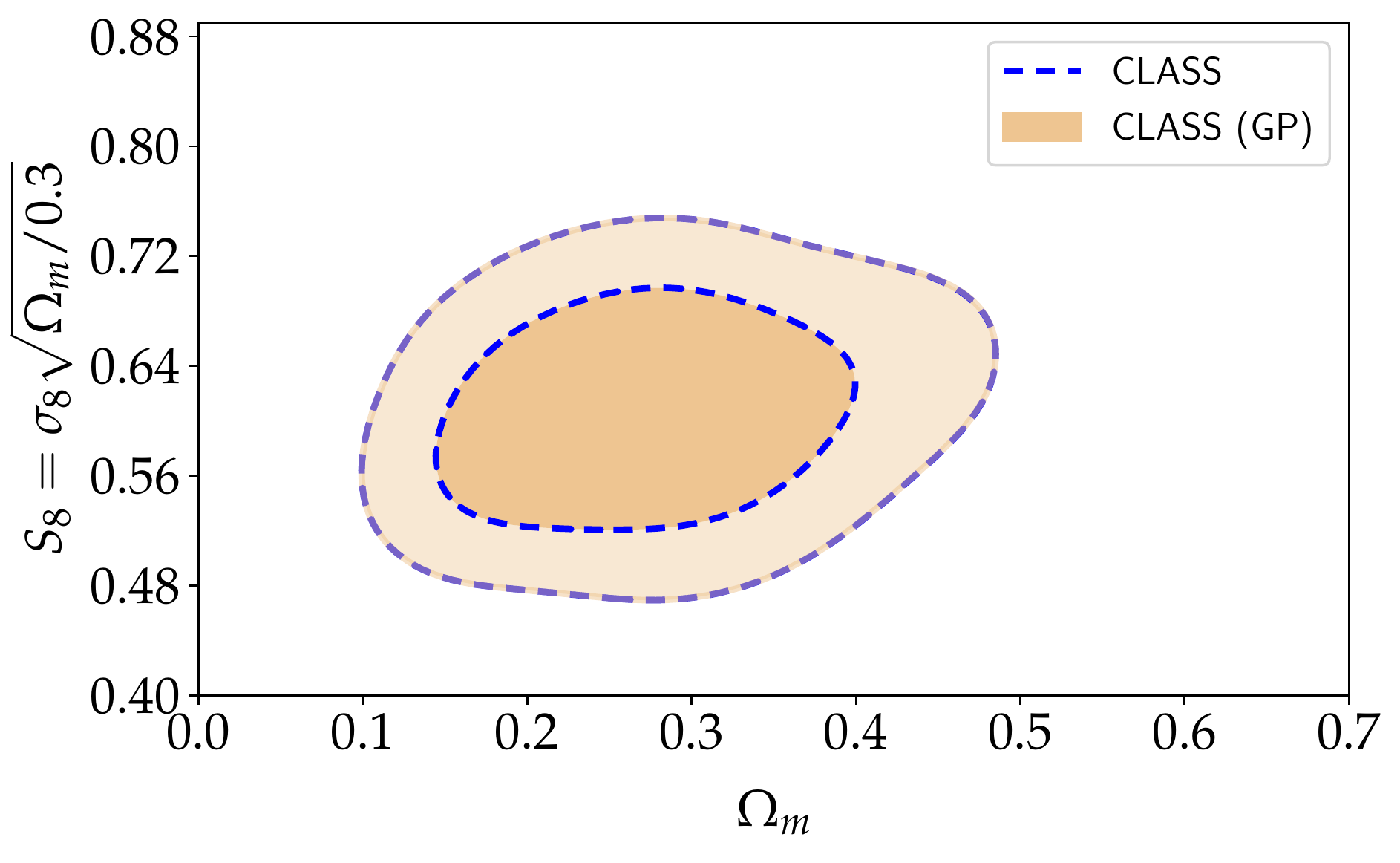}}}
    \qquad
    \subfloat[Without MOPED Compression]{{\includegraphics[width=0.30\textwidth]{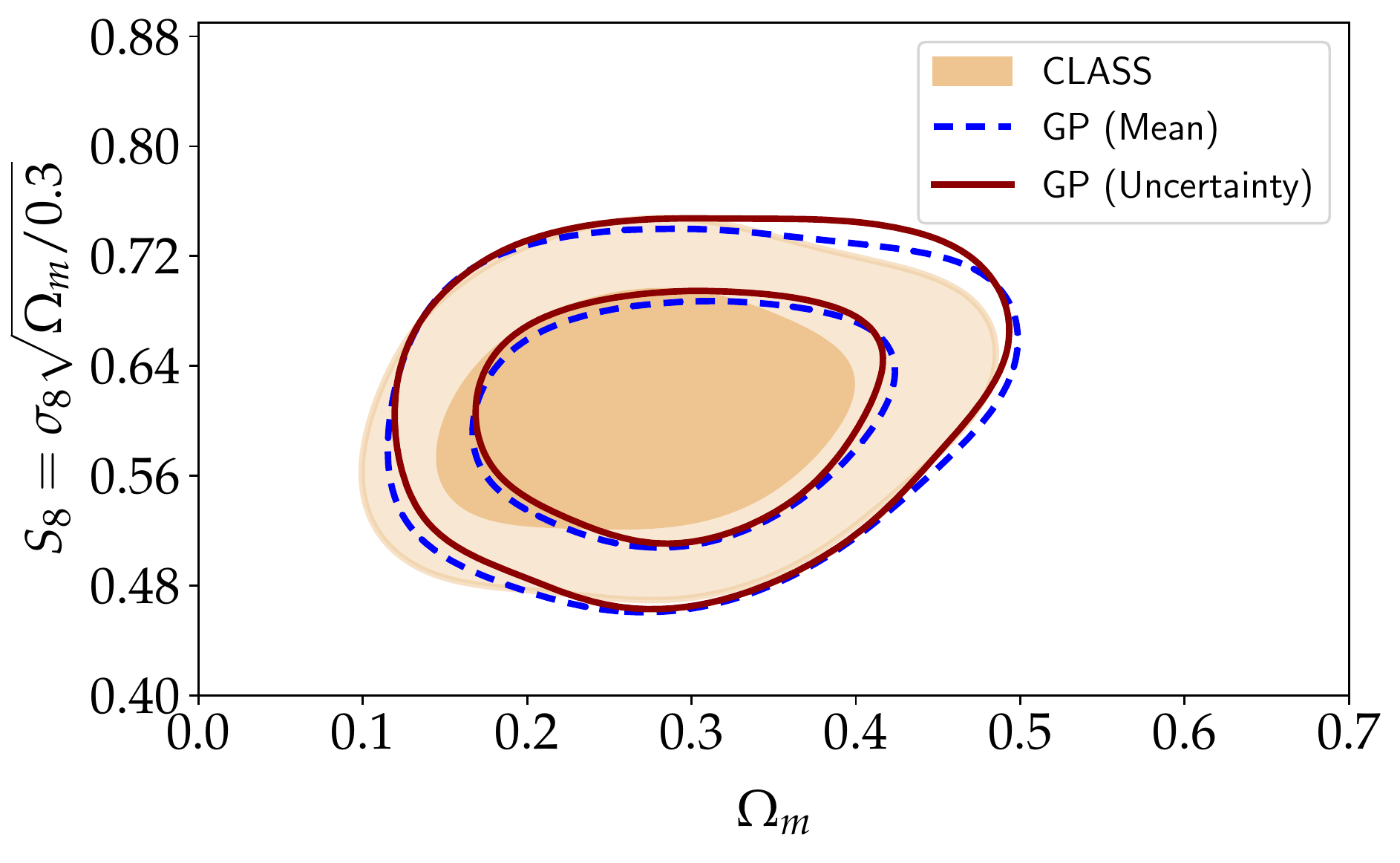}}}
    \qquad
    \subfloat[With MOPED Compression]{{\includegraphics[width=0.30\textwidth]{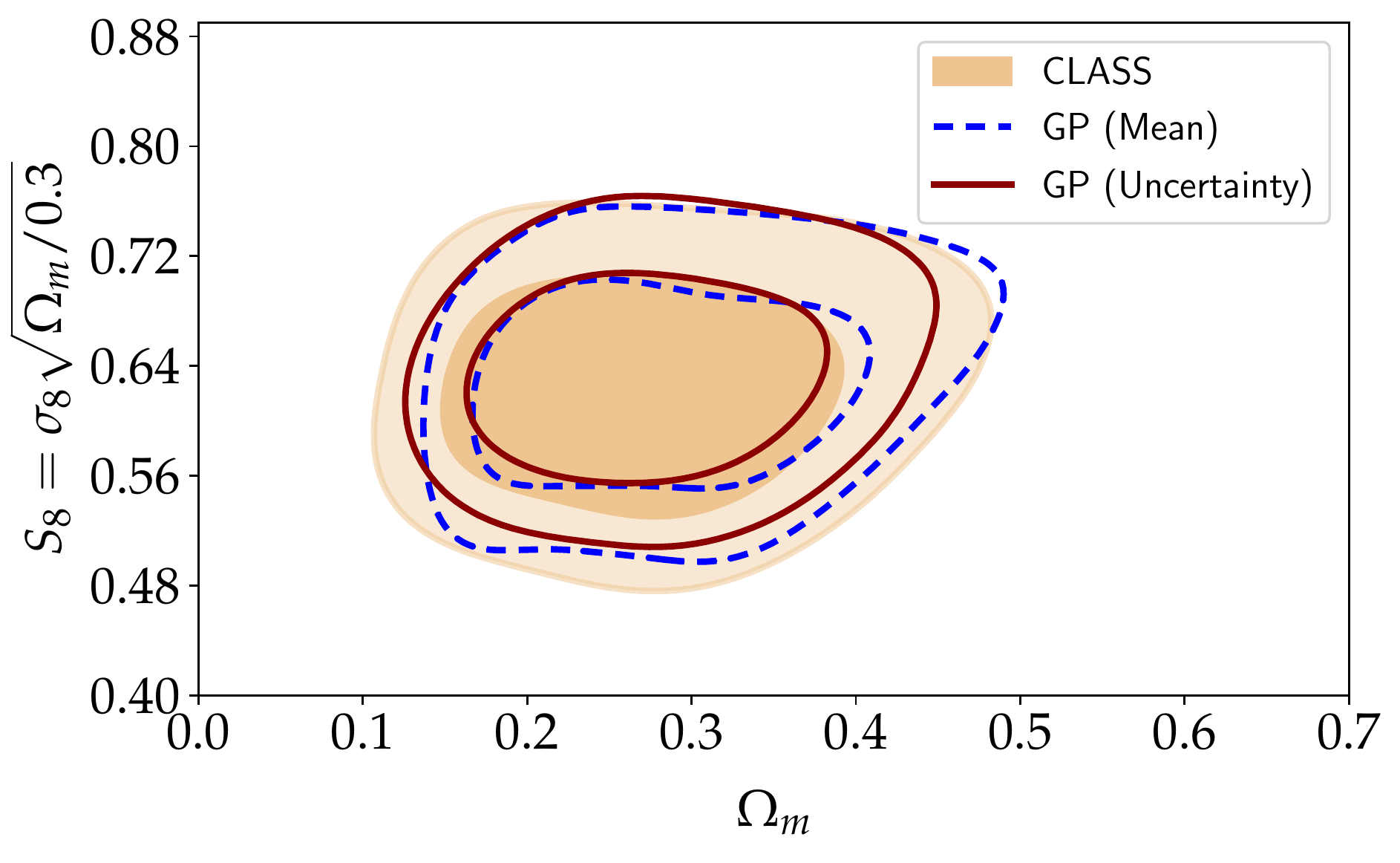}}}   
    \caption{\label{fig:s8_omega_matter}$S_{8}$ versus $\Omega_{m}$ plane for our analysis. The left panel shows that the Gaussian Process emulator, which is a function of our cosmological parameters, for computing $\sigma_{8}$ is accurate and precise enough compared to CLASS. The middle panel shows the constraints without MOPED compression while the right panel includes MOPED compression. The inner and outer contours correspond the 68\% and 95\% credible interval respectively.}
\end{figure*}

\noindent where $\bs{b}_{\alpha}^{\tm{T}} \bs{\mu}$ is computed using the emulator. The fact that the likelihood of the compressed data involves only $\mc{O}(p)$ operation makes parameter inference very fast since the $\mc{O}(N^{3})$ operation in the standard likelihood is completely eliminated, provided $\bs{b}^{\tm{T}}_{\alpha}\bs{\mu}$ can be rapidly computed.

By emulating the MOPED coefficients directly with separate Gaussian Processes, we have a very powerful tool. The GPs are still functions of just 8 parameters (6 cosmological and 2 systematics) and we now have only 11 separate GPs. Crucially, this setup is interesting because increasing the number of band powers (for example, in forthcoming lensing surveys) will not affect the MOPED timings at all.
\section{Results}
\label{sec:results}

Fig. \ref{fig:slice} shows 2 band powers, evaluated across the $\Omega_{\tm{cdm}}h^{2}$ slice in parameter space. In particular, the function in black corresponds to the accurate solver, CLASS while the broken red function corresponds to the GP mean, with the tan shading giving the $3\sigma$ credible interval of the GP. Note also that the right panel shows the GP prediction through a given training point and as expected, the GP uncertainty tends to zero. As seen in Fig. \ref{fig:slice}, the GP is able to predict the band powers quite well.

Since the predictive function is a Gaussian distribution, we can build a simple emulator by just using the mean, or propagate the uncertainty from the Gaussian Process through the model. Either method gives reasonable posterior densities as shown in Fig. \ref{fig:posterior_plot}. On a high end desktop computer, the evaluation is quite fast. Computing one likelihood with the mean of the Gaussian Process takes 0.03 seconds compared to 0.09 seconds if we include the Gaussian Process uncertainty with 20 Monte Carlo samples to marginalise over the GP uncertainty. On the other hand, CLASS takes 0.65 seconds for a likelihood evaluation. If we use 1000 training points, this yields an overall speed-up by a factor of $\sim12-30$ depending on whether we use the mean or the GP variance. In our case, we generate 360000 MCMC samples using EMCEE \citep{2010CAMCS...5...65G, 2013PASP..125..306F} for which the full simulator takes about 44 hours while the Gaussian Process emulator, using the mean, takes about $\sim 1.5$ hours. On the other hand, when we emulate the MOPED coefficients using 1000 training points, each likelihood computation takes $\sim 0.03$ seconds with either the mean or the variance of the GPs. As an example, with the MOPED compression, CLASS takes $\sim 44$ hours to generate 330000 MCMC samples (note that there is no significant improvement in speedup because we have just 24 band powers and each likelihood computation with or without the MOPED compression is almost the same). However, with the emulator, we obtain the same number of MCMC samples in $\sim 1.5$ hours with either the mean or variance of the GPs. All experiments with EMCEE were run on a single core. An interesting additional feature for the emulation scheme would be to exploit parallelization to speedup inference further.

The distribution of the log-posterior (up to a normalisation constant) of the MCMC samples obtained by using CLASS (in pale blue) is shown in Fig. \ref{fig:post_samples}. In the same plot, the red and green histograms show the distribution of the log-posterior when using the mean and error from the GP respectively. In the same figure in panel (b), we show the log-posterior of the samples obtained after compressing the data using the MOPED formalism. Note that, the distribution of the log-posterior of the different MCMC samples gives an indication of how faithful the function reconstruction with the GP is. With a small number of training points, there is a small shift of the log-posterior distribution of the GP emulator (either with the mean or the uncertainty) relative to the CLASS distribution.

\begin{table*}
\begin{minipage}{110mm}
\caption{\label{tab:results}Computational cost comparison between CLASS and the GP emulator}
\noindent \begin{centering}
\renewcommand\arraystretch{2.0}%
\begin{tabular}{cccccc}
\hline 
$N_{\tm{train}}$ & \textbf{Training} & \textbf{MCMC (Mean)} & \textbf{MCMC (Uncertainty)} & $D_{\tm{KL}}$ (Mean)& $D_{\tm{KL}}$ (Uncertainty)\tabularnewline
\hline 

1000 & 20 & 84 & 216 & 0.84 & 1.00 \tabularnewline
1500 & 48 & 85 & 290 & 0.63 & 0.89 \tabularnewline
2000 & 92 & 86 & 396 & 0.60 & 0.81 \tabularnewline
2500 & 139 & 88 & 524 & 0.47 & 0.68 \tabularnewline
3000 & 209 & 90 & 692 & 0.09 & 0.65 \tabularnewline

\hline 
\noalign{\smallskip} 
\multicolumn{6}{l}{\parbox{120mm}{\textit{Note}:The training and sampling time (columns 2,3,4) are given in minutes and the KL divergence is computed in units of nats (scaled by a constant); the largest $D_{\tm{KL}}$ is with the 1000 training points when we include the GP uncertainty, in which case, $D_{\tm{KL}}=2.03\times 10^{-12}$.}}\tabularnewline
\hline 
\end{tabular}
\par\end{centering}
\end{minipage}
\end{table*}

To compare the two distributions, we compute the Kullback-Leibler (KL) divergence between the CLASS distribution and the GP distribution, that is, 

\begin{equation}
D_{\tm{KL}}\left(p\left\Vert q\right.\right)=\sum\,p\left(\bs{\theta},\,\bs{\beta}\left|\bs{d}\right.\right)\,\tm{log}\left[\dfrac{p\left(\bs{\theta},\,\bs{\beta}\left|\bs{d}\right.\right)}{q\left(\bs{\theta},\,\bs{\beta}\left|\bs{d}\right.\right)}\right]
\end{equation}

\noindent where $p\left(\bs{\theta},\,\bs{\beta}\left|\bs{d}\right.\right)$ and $q\left(\bs{\theta},\,\bs{\beta}\left|\bs{d}\right.\right)$ are the posterior probabilities computed using CLASS and the GP at the same points in parameter space. Since the posterior probability is cheap to compute with the GP, we use all the MCMC samples obtained using CLASS to compute $q(\bs{\theta},\,\bs{\beta})$. The KL-divergence in nats, as a function of the number of training points, is shown in Table \ref{tab:results}. In general, the reconstruction of the band powers is almost perfect as the number of training points increases. This can also be deduced from the $5^{\tm{th}}$ column in Table \ref{tab:results} where the KL divergence decreases with increasing training points. If one could afford additional simulations, one option would be to just use the mean of the GP to sample the posterior distribution since it is not only faster compared to the case where the GP uncertainty is included, but is also closer to the actual true posterior distribution.

To assess the convergence of our MCMC chains, we also compute the Gelman-Rubin statistics \citep{1992StaSc...7..457G} for different scenarios. The latter is simply defined as $\hat{R} = \nicefrac{V}{W}$, where $V$ is the between-chain variance and $W$ is the within chain variance. $\hat{R}$ is calculated for different cases, for example, for a fixed number of training points, we use the MCMC samples using the GP (mean) and the MCMC samples obtained using CLASS. This is repeated with the MCMC samples where the GP uncertainty is included. In all cases, we apply a threshold of 1.05 to ensure that the chains satisfy the ergodicity condition.

We are also interested in the $S_{8}=\sigma_{8}\sqrt{\Omega_{\tm{m}}/0.3}$ cosmological parameter constraint. Recall that the GPs for sampling the posterior are built using the 8 parameters (6 cosmological and 2 systematics) and they do not allow us to predict $\sigma_{8}$ directly. However, the latter is a function of just the 6 cosmological parameters, since it involves an integration over the power spectrum. Therefore, as we compute the band powers with the 1000 training points, we also record the $\sigma_{8}$ values, as generated by CLASS. We then construct a training set with inputs

$$\left[\Omega_{\tm{cdm}}h^{2},\,\Omega_{\tm{b}}h^{2},\,\tm{ln}(10^{10}A_{\tm{s}}),\,n_{\tm{s}},\,h,\,\Sigma m_{\nu}\right]$$

\noindent which is then used to build an additional GP for $\sigma_{8}$. This then allows us to predict $\sigma_{8}$ at any point in the parameter space within the prior box. We find that it takes only 1 minute to predict $\sigma_{8}$ for 360 000 MCMC samples.   

In Fig. \ref{fig:s8_omega_matter}, we show the 2D marginalised posterior distribution of the derived parameters $S_{8}$ and $\Omega_{\tm{m}}$ using three different methods is shown. In particular, we compare the distribution obtained from CLASS with the mean and uncertainty of the GP and we conclude that we are able to recover comparable posterior densities for these two quantities, $S_{8}$ and $\Omega_{\tm{m}}$. 

In high dimensions, the GP uncertainty inflates between any two points. It is expected that adding more training points will improve the performance of the emulator (either with the mean or GP uncertainty) since the reconstruction of the emulated function will converge to the original function. In general, with increasing number of training points, the GP uncertainty will also decrease. The effect of the number of training points is indicated by the values of the KL-divergence in the last two columns of Table \ref{tab:results}. However, we empirically found that the KL-divergence when we use the mean of the emulator, is always better compared to the GP uncertainty. 
%This could be due to the fact that the GP uncertainty adds more stochasticity to the overall process in the likelihood computation.

One might expect the inclusion of the GP uncertainty to broaden the likelihoods, so the KL divergence would not be an appropriate measure of success. However, this does not appear to be the case: marginal errors are not noticeably increased. Our conclusion is that
inclusion of the GP uncertainty does not improve results, but this might vary with application. The reason is probably that we are emulating a precise function, where the training points have zero error, and  in this circumstance, the GP (which makes some assumptions that do not hold in detail) provides an error that is only approximately correct \citep{2020arXiv200110965K, 2020arXiv200201381W}.
\section{Conclusions}
\label{sec:conclusions}

\begin{figure}
\noindent \begin{centering}
\includegraphics[width=0.45\textwidth]{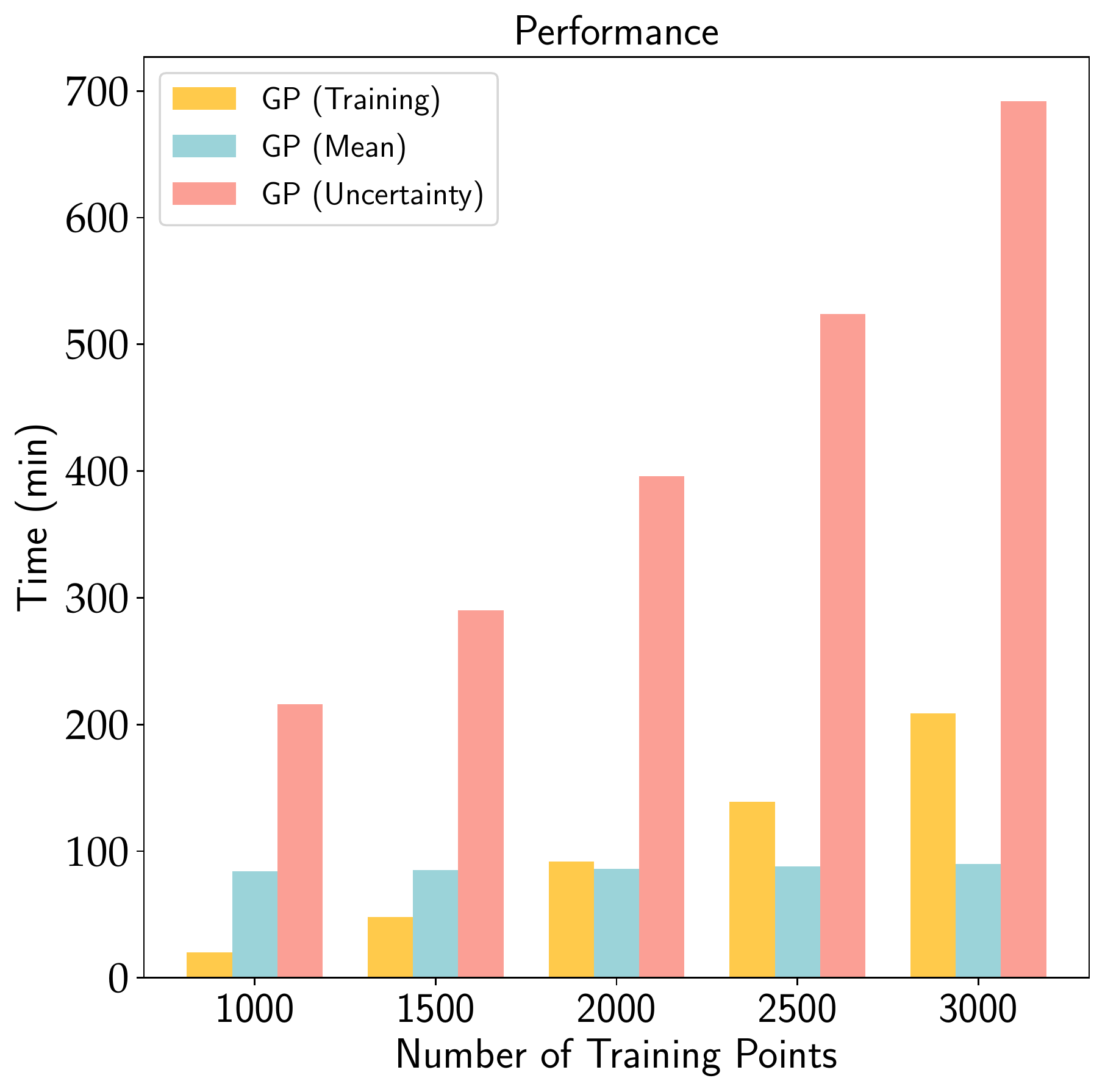}
\par\end{centering}
\caption{\label{fig:performance}Illustrating the performance of the emulator as a function of the number of training points. The expensive computations reside in training and predicting the GP uncertainty. Sampling the posterior with the GP mean is quick, even with the 3000 training points. The graphs do not perfectly follow the expected scaling with $N$ because of various overheads.}
\end{figure}

We have designed a principled and detailed Gaussian Process emulator for constraining not only weak lensing cosmological parameters but also the nuisance parameters. In summary, for this problem, 1) the (expensive) solver is queried a few thousand times only, to generate a training set (compared to a conventional MCMC routine where the solver is queried at every likelihood computation, 2) the emulator is $\sim 20$ times faster compared to the full solver and this makes inference very quick and 3) by emulating the MOPED coefficients, the number of separate Gaussian Processes is equal to the number of parameters in the model and inference, irrespective of the number of data points, and is a very powerful technique for analysing large datasets. Moreover, the posterior distributions obtained from the emulator are quite robust compared to the full run of the simulator, with and without MOPED. 

We have also demonstrated that the emulator can be used to emulate the MOPED coefficients directly. Both combined are expected to accelerate cosmological parameter inference. Emulating the MOPED compressed data has two major advantages. The first is a feature of MOPED itself, that the compressed data set does not grow at all as the original dataset increases in size, so scales exceptionally well to Euclid and LSST. The second is that MOPED is only fast if the theoretical values of the MOPED coefficients can be computed very quickly. The GP provides this functionality.  This is the most important conclusion of this paper.

In addition, we have used the KL-divergence as a metric to assess the performance of the emulator in obtaining reliable high dimensional posterior distributions. As evident from Table \ref{tab:results}, the larger the number of training points, the better the reconstruction of the emulated function and hence the lower the KL-divergence between the accurate CLASS posterior distribution and the emulator posterior distribution.

We also recommend using the mean of the emulator for this application. In Table \ref{tab:results}, the KL-divergence between the emulator posterior and the CLASS posterior shows that the mean is always better than the emulator with the GP uncertainty. From a computational perspective, this has various other advantages, for example, inference is very fast since the GP mean prediction requires $\mc{O}(N)$ operations (recall that the GP mean is a linear predictor) and storage.

An exciting application of this emulator can be in the case where one requires non-Limber computation of the power spectra. This certainly applies to galaxy clustering statistics \citep{2019arXiv191111947F} and the weak lensing bispectrum \citep{2020arXiv200401666D}, even if for the weak lensing power spectrum it is a good approximation \citep{2017PhRvD..95f3522K, 2017MNRAS.472.2126K}. In general, the latter is computationally expensive to be calculated accurately, especially at large $\ell$ because of the rapid oscillations of the spherical Bessel functions \citep{2017JCAP...05..014L}. For example, if the CLASS run were to be repeated without the Limber approximation, the emulator would have been $\sim 10^{3}$ times even faster. In future surveys, because the number of tomographics bins will be large, one would require more power spectra computations. For example, 10 tomographic bins lead to 55 auto- and cross- power spectra and the emulator would be $\sim 10^{3}$ and $\sim 10^{5}$ faster with and without the Limber approximation respectively.

Emulation has various other key advantages, apart from speeding up inference. As an example, one has to choose a good proposal distribution, which often requires tuning, to run an MCMC chain with the full simulator. The emulator can be used to explore the parameter space quickly and learn a suboptimal proposal distribution which can then be used with the full simulator. 

The accuracy of the reconstructed function can be improved by adding more training points as we have demonstrated. However, scaling Gaussian Processes to large number of training points results in a major computational bottleneck, mainly due to $\mc{O}(N^{3})$ operations in training and $\mc{O}(N^2)$ in predicting the uncertainty (see Fig. \ref{fig:performance}). Fortunately, here a few hundred training points suffices to give cosmological results with only a few percent degeneracy in error bars. Moreover, in this work, the training points have been placed according to the prior range itself. However, the interpolation scheme can be improved if we have more constrained parameters where we can use better prior information such as a Fisher matrix to intelligently place the training points. Alternatively, one can also do a quick optimisation to find the maximum likelihood estimator and the Hessian matrix, both of which can be used to construct an optimal design for the training points. 

An alternative option to accelerate the computation of GP uncertainty is to intelligently partition the training set by using a clustering algorithm, for example, $k-$means clustering \citep{hastie01statisticallearning}. During the prediction step, one can then use a local expert, which has a smaller kernel size, to compute the GP uncertainty swiftly. 

A quantity which is often required in optimisation or Monte Carlo methods such as Hamiltonian Monte Carlo (HMC) is the gradient with respect to the negative log-likelihood (cost function). Conveniently, the gradient with respect to the mean of the Gaussian Process surrogate model is analytic and this opens a new avenue towards accelerating gradient computation as well.

Gaussian Processes should not only be interpreted as a method for just accelerating computations. Instead, it effectively allows us to compute the posterior distribution of a function by placing a prior over it. In this work, the EE band powers and MOPED coefficients are modelled independently as Gaussian Processes and we have shown that we can recover robust cosmological parameters, whilst still marginalising over the nuisance parameters.

\section*{Acknowledgement}
We thank the referee for helpful and meaningful comments. AM is supported by the Imperial College President's Scholarship. We thank Jonathan Pritchard for suggesting the use of LH samples for building the emulator and Pat Scott, Daniel Jones for useful discussion. We also thank Zafiirah Hosenie for providing useful suggestions to improve this manuscript. We also thank Prof. Marc Deisonruth and Prof. David van Dyk for insightful discussion at the beginning of this project.

\section*{Data Availability}
The code and data products underlying this article will be made available at \href{https://github.com/Harry45/gp_emulator}{\texttt{https://github.com/Harry45/gp\_emulator}}.

%%%%%%%%%%%%%%%%%%%%%%%%%%%%%%%%%%%%%%%%%%%%%%%%%%
% \bibliographystyle{mnras}
\bibliographystyle{aa_url}
\bibliography{reference}

% Don't change these lines
\bsp	% typesetting comment
\label{lastpage}
\end{document}